\documentclass[prb,superscriptaddress,twocolumn,preprintnumbers,amsmath,amssymb]{revtex4-2}
  
\usepackage[breaklinks,colorlinks=true]{hyperref}
\usepackage{mathtools}
\usepackage{slashed}
\usepackage{tensor}
\usepackage{bbold}
\usepackage{amscd}
\usepackage{calrsfs}
\DeclareMathAlphabet{\pazocal}{OMS}{zplm}{m}{n}
\allowdisplaybreaks

\newcommand{\cgamma}{\underline{\gamma}\vphantom{\gamma}}

\begin{document}
\title{Curvature-induced pseudogauge fields from time-dependent geometries in graphene}
\author{Pablo A. Morales}
\email{pablo$_$morales@araya.org}
\affiliation{Research Division, Araya Inc., Tokyo 107-6019, Japan}
\affiliation{Centre for Complexity Science, Imperial College London, London SW7 2AZ, UK}
\author{Patrick Copinger}
\email{copinger0@gate.sinica.edu.tw}
\affiliation{Institute of Physics, Academia Sinica, Taipei 11529, Taiwan}

\begin{abstract}
The massless Dirac equation is studied in curved space-time on the (2+1) dimensional graphene sheet in time-dependent geometries. Emergent pseudogauge fields are found both in the adiabatic regime and, for high-frequency periodic geometries, in the nonadiabatic regime for a generic Friedmann–Lemaître–Robertson–Walker metric in Fermi normal coordinates. The former extends the conventionally understood homogeneous pseudogauge field to include weak temporal inhomogeneities. The latter, through the usage of Floquet theory, represents a new class of emergent pseudogauge field, and is argued to potentially provide a condensed matter realization of cosmological high-frequency geometries.
\end{abstract}

\maketitle

\section{Introduction}

The realization of a Dirac-like dispersion relation in graphene~\cite{Novoselov2005,Geim2007} has made possible an interdisciplinary bridge between condensed matter, high-energy physics, and gravity. In particular, in curved space the study of the Hawking effect~\cite{IORIO2012334,PhysRevD.90.025006}, BTZ black holes~\cite{CVETIC20122617}, and even wormholes~\cite{GONZALEZ2010426}, among others, has been made possible through corrugations in graphene. A striking example is provided through the appearance of strain-induced pseudo-magnetic fields~\cite{PhysRevLett.97.016801,*PhysRevLett.97.196804}, observed even with previously inaccessible strength in excess of 300 T~\cite{doi:10.1126/science.1191700}. pseudogauge fields in graphene have been widely discussed~\cite{PhysRevB.76.165409,*Vozmediano_2008,*DEJUAN2010625,*PhysRevB.87.165131} and moreover are thought relevant for fundamental quantum field theory (QFT) descriptions of high-energy phenomena. For example, SU(2) monopole like gauge fields were observed from the intrinsic curvature of fullerene molecules~\cite{PhysRevLett.69.172,gonzalez1993electronic} and, furthermore, Abelian and non-Abelian, with Wu-Yang ambiguity, pseudogauge field differences from elastic degrees of freedom were explored in~\cite{PhysRevB.87.125419}. Let us also stress that pseudogauge fields are not limited to graphene, e.g., Weyl semimetals~\cite{PhysRevLett.115.177202} and their pseudogauge manifestation from its torsional strain have been well studied~\cite{Soto-Garrido_2018,*PhysRevResearch.2.012043,*nano12203711}. For a review of pseudogauge fields in graphene see~\cite{VOZMEDIANO2010109}.

A limitation of our current understanding of pseudogauge fields in graphene is that the emergent fields are modeled as homogeneous. While this approach is important for physical opacity and to make manifest the experimental realization of, e.g., Landau levels of the pseudomagnetic field~\cite{doi:10.1126/science.1191700} (whose emergence was also argued in a quasistatic nanoscale motor in graphene in~\cite{PhysRevE.91.052152}), it is important to extend our understanding to the more realistic time-dependent curvature induced fields. Furthermore, many curved space backgrounds of interest are time-dependent, and thus for condensed matter realizations it is essential that we explore the more general time-dependent case. Hence our aim in this work is to provide a map from time-dependent curved space geometries to their corresponding quantum mechanical system. To achieve this we explore two cases.
\begin{enumerate}
    \item The adiabatic case --- here the bispinor Green's function approach (which has been utilized successfully in the homogeneous case~\cite{pseudoBgraphene}) is extended to encompass adiabatic time-dependent geometries. A phenomenological merit is that existing experimental setups good for the homogeneous case may readily observe new adiabatic features. 
    \item The nonadiabatic and periodic with high-frequency case --- here we explore a novel emergent field structure in graphene using Floquet theory~\cite{oka2019floquet}, the merit of which here, it is argued, is that with application of an engineered high-frequency time-dependent magnetic field onto graphene, the corresponding curved space analog system may be realized on a tabletop experiment. This is thought important for high-frequency setups such as are found for rotating neutron stars, compact binaries, and early universe quantum fluctuations, among others.
\end{enumerate}

To concretely study time-dependent geometries in graphene we examine the case of a Friedmann–Lemaître–Robertson–Walker (FLRW) metric in (2+1) dimensions. This is specifically examined because its simple yet general profile may readily be modeled (for shorter distances than the Hubble radius~\cite{PhysRevD.79.064036}) on the graphene sheet for either weakly inhomogeneous times or in high-frequency Floquet induced systems.

Furthermore, the FLRW geometry covers a number of cosmological setups of interest. Floquet techniques have been used in the study of preheating scenarios following inflation via parametric resonance~\cite{allahverdi2010reheating}. Importantly, the idea that fermionic matter on FLRW systems plays an important role in cosmology has been gaining increasing attraction. For example, in~\cite{InflationFromFerm}, a curvature dependent fermionic mass inspired by covariant Hamiltonian approaches has been reported to not only successfully drive inflation, but also to produce a tensor-to-scalar ratio in agreement with observations among other features. In addition, lepton number asymmetry induced by fermionic matter during inflationary epoch is being actively considered as an alternative mechanism of baryogenesis~\cite{AxionInflation}. Last, spontaneous $\pazocal{CP}$ violation may generate dark fermions in inflation~\cite{maleknejad2020dark}.

To produce a map to a corresponding quantum mechanical system one can make use of a coordinate system that is locally flat. In contrast to time-independent fields where the connection to a local Lorentz frame may be achieved with Riemann Normal Coordinates (RNC), here for the general time-dependent case and proper interpretation of an observable in curved space~\cite{ParkerLinearH,ParkerPRL} one requires a frame along the worldline, and thus usage of Fermi Normal Coordinates (FNC)~\cite{fermi1921sull,Marzlin:1994wc,Nesterov_1999} is indispensable. 

This work is organized as follows: In Sec.~\ref{sec:chiral_fermions} we provide our setup of the Dirac equation in curved space motivating FNC. In Sec.~\ref{sec:HamL} we argue an effective Hermitian Hamiltonian based on the scalar product making connection to a local flat space. Then we illustrate through a more conventional approach, namely with the bispinor Green's function in Sec.~\ref{sec:QHam}, how emergent fields may be identified adiabatically; we concretely examine the case of an FLRW background in Sec.~\ref{sec:FermiCoords}. Finally, we argue the identification of a new class of emergent fields for the unique case of a Floquet engineered graphene in Sec.~\ref{sec:Floquet}.

For our notations, we use Greek symbols for curved indices, and for flat Lorentz indices we use the Latin letter set $\{a,b,c,d\}$. Other Latin letters are reserved for spatial indices. We use a mostly plus flat space metric denoted as $\eta_{ab}$. Our spin tensor in flat space reads $\Sigma^{ab}=[\gamma^{a},\gamma^{b}]$. In (2+1) dimensions our gamma matrices are $\gamma^{0}=-i\sigma^{3}$ and $\gamma^{i}\gamma^{0}=\sigma^{i}$ for $i=1,2$; $\sigma^a$ being the ordinary Pauli matrices.

\section{Chiral Fermions in curved space-time}
\label{sec:chiral_fermions}

The scale that governs the graphene dynamics, the Fermi velocity, is a hundredth of the speed of light bringing physics to relativistic domain and the particle-hole system is describable with chiral fermions. Due to the $\sigma$-bonds formed by the carbon atoms in the graphene monolayer, the sheet is able to withstand up to $25\%$, in elastic strains~\cite{cao2020elastic,*lee2008measurement}.
Last, out-of-plane ripples may result from topological defects such as occasional pentagon/heptagon dislocations in the honeycomb lattice. These properties allow one to model the graphene system as quasiparticles subject to a curved space-time.
Therefore, we make use of a curved QFT approach to graphene. Such an approach has proved indispensable to graphene physics leading to experimental discoveries such as including the pseudo-quantum Hall effect~\cite{Novoselov2005,Zhang2005}, space-time dependent Fermi velocities~\cite{PhysRevB.87.075405}, and the Klein paradox~\cite{Katsnelson2006}.

The dynamics of the collective modes on graphene are governed by the massless Dirac equation~\footnote{Let us point out that if one were to model intrinsic (membrane) properties rather than the extrinsic (embedding) in graphene then it has been advocated to use both Dirac points, rather the single one assumed here~\cite{OLIVALEYVA20152645,AntiIorio}}. Let us review some basics of the dynamics of chiral fermions on curved space. We will perform most of the analysis in d+1 dimensional space-time only to specialize on the graphene sheet, (2+1) dimensions, later on.
\begin{equation}
    \label{eq:Dirac_eq}
    i \underline{\gamma}^{\mu}(x)\nabla_{\mu} \psi = 0\,,
\end{equation}
where the $\cgamma$'s
denote space-time dependent $\gamma$-matrices related to the usual $\gamma$ matrices as $\cgamma_{\mu}(x) = \gamma_{a} e^{a}_{\hphantom{a}\mu}(x)$, where $e^{a}_{\hphantom{a}\mu}(x)$ are vielbein vector fields describing the metric, 
$g_{\mu \nu}(x) = e^{a}_{\hphantom{a}\mu}(x)e^{b}_{\hphantom{b}\nu}(x) \eta_{a b}$.
$\{a,b,c,d\}$ correspond to flat Lorentz indices.
The spinorial affine connections are defined by the vanishing of the covariant derivative of $\cgamma_\mu$,
that is,
\begin{equation}
    \nabla_{\mu}\cgamma_{\nu}=\partial_{\mu}\cgamma_{\nu} - \Gamma^{\lambda}_{\mu \nu}\cgamma_{\lambda} - [\Omega_{\mu}, \cgamma_{\nu}] = 0\,.
\end{equation}
The covariant derivative acting on the spinor fields with spin connection, $\Omega_\mu$, is
\begin{equation}
    \nabla_{\mu} \psi =(\partial_{\mu} - \Omega_{\mu}) \psi = \Bigl(\partial_{\mu} - \frac{1}{8}\omega_{\mu}^{a b}\Sigma_{a b}\Bigr) \psi \,,
\end{equation}
where the following are understood:
\begin{subequations}
\begin{align}
    \Sigma^{ab} & = [\gamma^{a},\gamma^{b}]\,, \\
    \omega_{\mu}^{a b} & = e^{a}_{\nu}g^{\nu\lambda}(\Gamma^{\sigma}_{\mu \lambda} e^{b}_{\sigma} - \partial_{\mu} e^{b}_{\lambda})\,.
\end{align}
\end{subequations}

\begin{figure}
  \includegraphics[width=8.5cm,height=6cm,keepaspectratio]{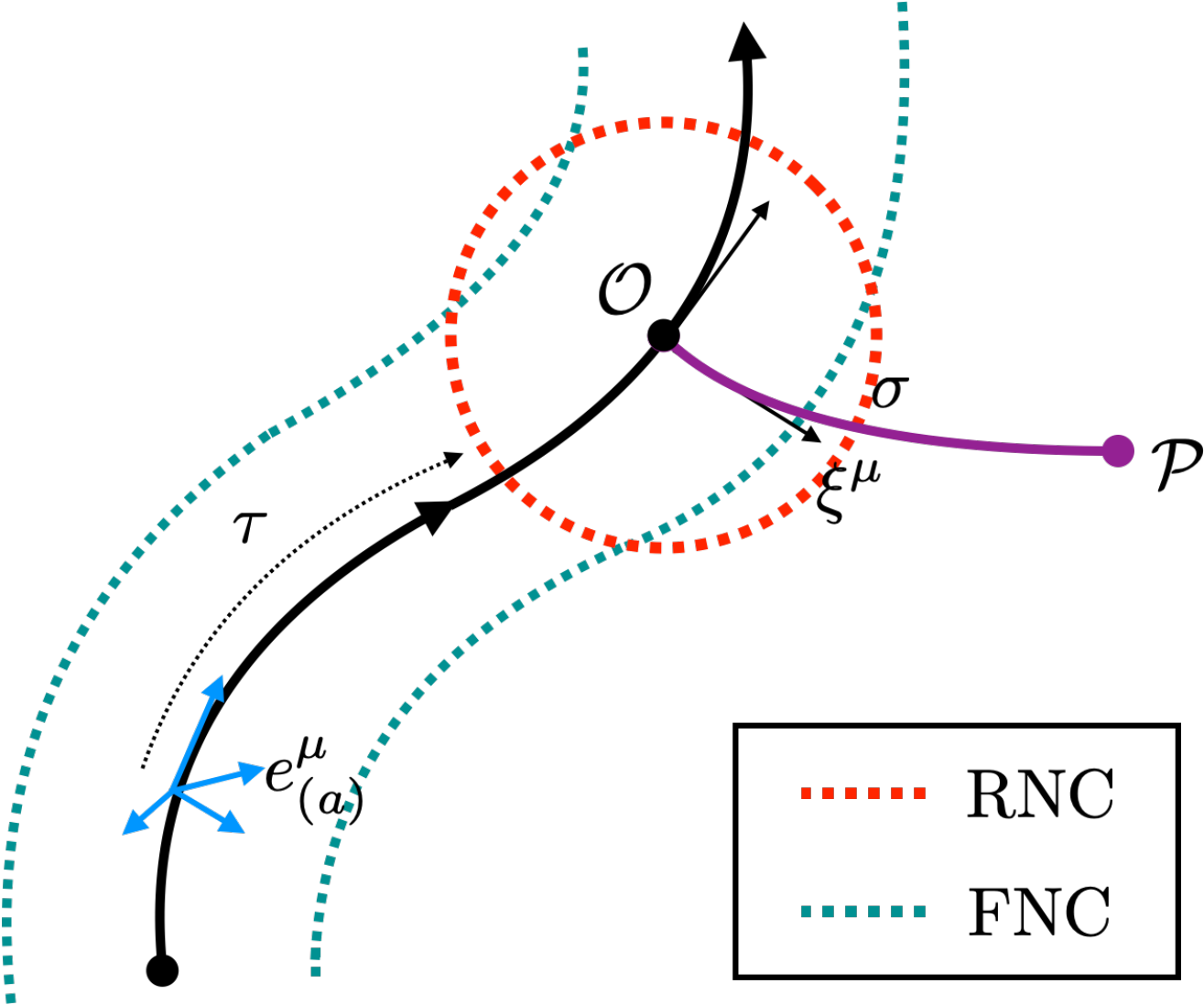}
  \caption{Trajectory with geodesic with proper time, $\tau$, and observer $\pazocal{O}$ is described by means of FNC with corresponding cylindrical region of validity, extending the notion of the its cousin, RNC with spherical region of validity only in the neighborhood of $\pazocal{O}$. For FNC, $y^\mu$, the timelike vielbein, $e^0_{(a)}$, is tangent to the trajectory's geodesic, and for given proper time an orthogonal hypersurface extending to $\pazocal{P}$ by proper distance $\sigma$ may be constructed such that $y^\mu=(\tau,\sigma \xi^\nu e_\nu^{(i)})$~\cite{manasse1963fermi,*dSFermiCoords}.}
  \label{fig:fnc}
\end{figure}

To interpret an emergent quantum mechanical framework, our next task is to determine a suitable coordinate system that is locally flat. Usage of RNC is successful in this regard for time-independent systems. To study general emergent phenomena, such as is required for high frequency Hamiltonians found in Floquet engineered~\cite{oka2019floquet} graphene and semimetals, we must treat time dependence. To this end we make use of FNC~\cite{fermi1921sull}; see also~\cite{Marzlin:1994wc,Nesterov_1999}. FNC have proved valuable in the determination of energy-levels of a single atom in curved space-time~\cite{ParkerLinearH,ParkerPRL}. This is because FNC supplies one with a continuous locally inertial reference frame along the observer's worldline making possible the determination of a well-defined observable. In essence FNC are defined by the Fermi conditions; $g_{\mu \nu}|_G = \eta_{\mu \nu}$ and $\Gamma_{\mu \hphantom{\rho} \nu}^{\hphantom{\mu} \rho}|_G = 0$ imposed along the geodesic $G$. With time $y^0=\tau$ and proper time $\tau$, one may construct an orthogonal set of timelike and spacelike geodesics, to which the timelike geodesic runs parallel to proper time and the spacelike hypersurface intersects it; see Fig.~\ref{fig:fnc}. In this way the region of validity of the coordinates form a cylinder about the proper time; this is in contrast to RNC, whose validity only exists at a single space-time point (taken as $\pazocal{O}$ in Fig.~\ref{fig:fnc}) and are thus inadequate for time-dependent emergent systems from curved space. To specifically construct FNC, one may examine a second geodesic, which is orthogonal to the geodesic traversed by $\tau$ and thus serving as a spacelike hypersurface, extending a proper distance $\sigma$ away from point $\pazocal{O}$ to point $\pazocal{P}$. Then one may find in a rigorous way FNC such that $y^\mu=(\tau,\sigma \xi^\nu e_\nu^{(i)})$ for vielbein $e_\nu^{(i)}$. See Refs.~\cite{manasse1963fermi,*dSFermiCoords} for further details.

The metric in FNC reads~\cite{manasse1963fermi}
\begin{subequations}\label{eq:fnc_metric}
\begin{align}
    g_{00} &= -1 - R_{0l0m}y^{l}y^{m} + \cdots \,, \\
    g_{0i} &= -\frac{2}{3} R_{0lim}y^{l}y^{m} + \cdots\,, \\
    g_{ij} &= \delta_{ij} - \frac{1}{3} R_{iljm}y^{l}y^{m} + \cdots\,,
\end{align}
\end{subequations}
where Latin indices, with exception to $\{a,b,c,d\}$, denote space components. Here $R_{ijlm}$ are to be understood as curvature components evaluated on the geodesic, with $R_{ijlm}|_G$ omitting the evaluation on $G$ for notation brevity. The inverses can be easily read off from the above expression by a change of sign of the second term of $g_{00}$ and $g_{ij}$, whereas $g^{0i}$ remains unchanged. With the above one may determine
\begin{equation}
    g=-1+\frac{1}{3}(R_{lm}-2R_{0l0m})y^{l}y^{m} + \cdots \,,
\end{equation}
where $g\coloneqq \mathrm{det}(g_{\mu\nu})$. The time dependence of this expressions is completely contained within the curvature components whereas space dependence is explicitly shown in the expansion, in contrast to RNC where the expansion is considered around a fixed space-time point. For our purposes the first term of the expansion suffices to address spatial dependence. In a similar way to $g$, the vielbein fields are
\begin{subequations}\label{eq:vielbein}
\begin{align}
    e^{a}_{\hphantom{a}0} &= \delta^{a}_{0} - \frac{1}{2}R^{a}_{\hphantom{a}l0m}y^{l}y^{m}\,, \\
    e^{a}_{\hphantom{a}i} &= \delta^{a}_{i} - \frac{1}{6}R^{a}_{\hphantom{a}lim}y^{l}y^{m}\,.
\end{align}
\end{subequations}
With the above we can write in FNC the Christoffel connections to linear order in curvature, whose components read
\begin{subequations}
\label{eq:Cconn}
\begin{align}
    \Gamma_{00}^{0}&=\frac{1}{2}\partial_{0}R_{0l0m} y^{l}y^{m}\,,\\
    \Gamma_{0i}^{0}&=R_{0i0m}y^{m}\,,\\
    \Gamma_{00}^{i}&=R_{0i0m}y^{m} -\frac{2}{3} \partial_{0}R_{0lim}y^ly^m\,,\\
    \Gamma_{ij}^{0}&=\frac{1}{3}(R_{0ijm}+R_{0jim})y^{m} -\frac{1}{6}\partial_{0}R_{iljm}y^{l}y^{m}\,,\\
    \Gamma_{0j}^{i}&=R_{0mji}y^{m} - \frac{1}{6}\partial_{0}R_{iljm} y^{l}y^{m}\,,\\
    \Gamma_{jk}^{i}&=\frac{1}{3}(R_{jikm}+R_{kijm})y^{m}\,.
\end{align}
\end{subequations}
The above differ with those obtained in~\cite{ParkerLinearH} in which terms with time derivatives were dropped under the criteria of temporal slowly varying fields. For the analysis of time-dependent background geometries, we employ FNC which is valid for arbitrary time. Spatial deformations in FNC are addressed by the inclusion of higher order terms in the $y$-expansion as needed. 
Here we keep the discussion within the weak regime, hence keep terms up to $\pazocal{O}(y_i^2)$ implying linear curvature. 
With Eq.~\eqref{eq:Cconn} one can determine the components of the spin connection as
\begin{subequations} \label{eq:SpinC_FNC}
\begin{align}
    \Omega_{0} &= \frac{1}{2} \gamma_{0}\gamma_{j} R^{j}_{\hphantom{j}00m}y^{m} + \frac{1}{4} \gamma_{k}\gamma_{j} R^{kj}_{\hphantom{kj}0m}y^{m} \nonumber \\
    &\hphantom{=}\;\, + \frac{1}{12} \gamma_{0}\gamma_{j} \partial_{0}R^{j}_{\hphantom{j}l0m}y^{l}y^{m}\,, \\
    \Omega_{i} &= \frac{1}{4} \gamma_{0}\gamma_{j} R^{0j}_{\hphantom{0j}im}y^{m} + \frac{1}{8} \gamma_{k}\gamma_{j} R^{kj}_{\hphantom{kj}im}y^{m} \nonumber \\
    &\hphantom{=}\;\, + \frac{1}{12} \gamma_{0}\gamma_{j} \partial_{0}R^{j}_{\hphantom{j}lim}y^{l}y^{m}\,.
\end{align}
\end{subequations}

Equipped with the above, we are now in a position to explore effective and emergent systems made possible with FNC. We first analyze a linear effective Hamiltonian construction by means of the scalar inner product, that will lead us to a high-frequency description for usage in a Floquet environment. Then we analyze a quadratic form that emphasizes certain emergent features in an adiabatic sense.

\section{Effective Hermitian Hamiltonian}
\label{sec:HamL}

Let us begin our discussion of an effective Hamiltonian by examining the massless Dirac equation, Eq.~\eqref{eq:Dirac_eq}. Such a Hamiltonian has been well studied in the context of a one-electron atom's spectrum~\cite{ParkerLinearH,ParkerPRL}, in condensed matter analog systems with gapless dispersion, among others. Rearranging terms, one can recast the massless Dirac equation as a Schr\"{o}dinger-like equation, i.e, $i\partial_0 \psi = \pazocal{H} \psi$, whose Hamiltonian reads~\cite{ParkerLinearH},
\begin{equation}\label{eq:Lin_Ham}
    \pazocal{H} = -i(g^{00})^{-1}\cgamma^0 \cgamma^i \nabla_i +i\Omega_0\,.
\end{equation}
Let us adopt FNC according to Eqs.~\eqref{eq:fnc_metric}-\eqref{eq:vielbein}. Then one may find
\begin{align}
    \pazocal{H} &= -i\gamma_{0}\gamma^{i}(\partial_i -\Omega_i) + i \Omega_0 \nonumber \\ & \hphantom{=}\;\, -\frac{i}{2}y^{l}y^{m}(R_{0l0m}\gamma_0 \gamma^i \partial_i + R_{0lim}\partial^i) \nonumber \\ 
    & \hphantom{=}\;\, - \frac{i}{6}y^{l}y^{m}(R_{iljm}\gamma_0 \gamma^j \partial^i + R_{0ljm}\gamma^j \gamma^i\partial_i)\,.
    \label{eq:H_F_FNC}
\end{align}
By virtue of the FNC for a time-dependent system we find the massless Dirac Hamiltonian is expressible with flat indices. 

An immediate drawback of Eq.~\eqref{eq:H_F_FNC} can be seen in that non-Hermitian terms are present, making comparison to an emergent quantum mechanical setting challenging. Indeed particle number conservation is not conserved in a time-dependent metric. However, for the purpose of constructing an emergent quantum mechanical Hamiltonian, $h_\text{eff}$, as an observable with associated well-defined real energy spectrum, let us confine our attention to the single particle Hamiltonian with conserved and Hermitian scalar product. 

\citet{ParkerLinearH,ParkerPRL}, and further by~\citet{PhysRevD.79.024020}, have demonstrated that a Hamiltonian may be constructed that is Hermitian with respect to the \textit{curved} scalar product,
\begin{equation}\label{eq:scalar_product}
    (\phi,\mathcal{H}\psi)\coloneqq-\int d^{3}x\sqrt{-g}\phi^{\dagger}\gamma^{0}\cgamma^{0}(x)\mathcal{H}\psi\,,
\end{equation}
such that $(\phi,\mathcal{H}\psi)=(\mathcal{H}\phi,\psi)$ for Hermicity. $\mathcal{H}$ is the Dirac Hamiltonian, Eq.~\eqref{eq:Lin_Ham}, augmented to ensure Hermicity and reads
\begin{equation}
    \mathcal{H} \coloneqq  \pazocal{H}- \Delta\pazocal{H}\,,
\end{equation}
with Hermitian correction factor
\begin{equation}
    \Delta\pazocal{H}=\frac{i}{2}\frac{\cgamma^{0}\gamma^{0}}{g^{00}\sqrt{-g}}\partial_0[\sqrt{-g}\gamma^{0}\cgamma^{0}]\,.
\end{equation}
Let us now demonstrate the role of the above correction, again making use of FNC. One can find to leading order in curvature in FNC the corrective factor becomes
\begin{equation}
    \Delta\pazocal{H}=\frac{i}{12}(\partial_{0}R_{0l0m}+\partial_{0}R_{lm}-\partial_{0}R_{0lim}\gamma^{0}\gamma^{i})y^{l}y^{m}\,.
\end{equation}
$\Delta\pazocal{H}$ has the effect of canceling out the time derivative contributions from the spin connection in Eq.~\eqref{eq:SpinC_FNC}, which would otherwise, break the Hermicity of the Hamiltonian in Eq.~\eqref{eq:Lin_Ham}. The second and third term cancel the time derivative at $\Omega_i$, whereas the first cancels out with that of $\Omega_0$.

To define our effective quantum mechanical Hamiltonian, let us take the scalar product in Eq.~\eqref{eq:scalar_product}, however expressed in flat space indices, and thus depicting a \textit{flat} scalar product. In this way, one may make sense of the spectrum and expectation values of observables. We take terms emerging from the deformed scalar product as contributions of an effective potential to the Dirac Hamiltonian describing our system. To arrive at the effective Hamiltonian let us note that the equivalent quantum mechanical scalar product reads, with single particle Hamiltonian,
$h_{\text{eff}}$, as
\begin{equation}\label{eq:flat_scalarprod}
    (\phi,h_{\text{eff}}\psi)_{0}\coloneqq\int d^{3}x\,\phi^{\dagger}h_{\text{eff}}\psi\,.
\end{equation}
Equating the above with the curved space definition in Eq.~\eqref{eq:scalar_product}, we can arrive at the corresponding effective quantum mechanical Hamiltonian,
\begin{equation}
  h_{\text{eff}}\coloneqq-\sqrt{-g}\gamma^{0}\cgamma^{0}(x)\mathcal{H}_{F}\,.  
\end{equation}
In FNC the additional factor results in an overall correction:
\begin{equation}
    h_{\text{eff}}=\Bigl\{1-\frac{1}{6}[R_{lm}+R_{0l0m}+R_{il0m}\gamma^{i}\gamma^{0}]y^{l}y^{m}\Bigr\}\mathcal{H}_{F}\,.
\end{equation}
And keeping to first order in curvature we finally find
\begin{align}\label{eq:effec_ham}
    h_{\text{eff}} &= -i\gamma^i\gamma^0 (\partial_i -\bar{\Omega}_i) + i \bar{\Omega}_0 + \frac{i}{6}y^{l}y^{m}[R_{lm}\gamma_i\gamma^0 \nonumber \\
    &\hphantom{=}\;\, + 2R_{l00m}\gamma_i\gamma^0 - R_{iljm}\gamma^j\gamma^0 - 3R_{0lim}]\partial^{i} \,.
\end{align}
The modified spin connections read
\begin{subequations}
\begin{align}
    \bar{\Omega}_{0} &= \Omega_0 - \frac{1}{12} \gamma_{0}\gamma_{j} \partial_{0}R^{j}_{\hphantom{j}l0m}y^{l}y^{m}\,, \\
    \bar{\Omega}_{i} &=\Omega_i -\frac{1}{12} \gamma_{0}\gamma_{j} \partial_{0}R^{j}_{\hphantom{j}lim}y^{l}y^{m}\,.
\end{align}
\end{subequations}
with the spin connections given in Eq.~\eqref{eq:SpinC_FNC}. The above Hamiltonian is Hermitian, and let us simply show that is indeed the case. One should notice the offensive terms in the spin connections, i.e., the first term of $\Omega_0$ and the first term of $\Omega_i$ in~\eqref{eq:SpinC_FNC}, which both fail to preserve Hermicity. However additional terms in~\eqref{eq:effec_ham} yield new contributions from the derivative operator, which are
\begin{align}\label{eq:hermit_corr}
    h_{\text{eff}}^{\dagger} \ni \frac{i}{6}(\delta^{l}_{i}y^m + y^l \delta^{m}_{i}) &[\gamma^i\gamma^0(R_{lm}+ 2R_{l00m}) \nonumber \\
    & - \gamma^j\gamma^0 R_{iljm} - 3R_{0lim}] \,.
\end{align}
Their addition has the effect of ultimately ensuring the Hamiltonian is Hermitian, $h_{\text{eff}}^{\dagger} = h_{\text{eff}}$. Eq.~\eqref{eq:effec_ham} is valid for arbitrary time dependence including time derivatives of Riemann curvature, stemming from its curved space analog by virtue of FNC. 

We have formulated an emergent and effective Hamiltonian and we would like to briefly turn our attention to specific terms that resemble emergent fields. To do so, let us turn to the general case in the (2+1) dimensional graphene sheet of arbitrary geometry, in which the spin connections become 
\begin{subequations}
\label{eq:spinconn_FNCgen}
\begin{align}
    \bar{\Omega}_{0}&=\frac{1}{2}\sigma^{j}R_{j00m}y^{m}-\frac{i}{2}\sigma^{3}R_{12m0}y^{m}\,, \\
    \bar{\Omega}_{i}&=\frac{1}{4}\sigma^{j}R_{0jim}y^{m}-\frac{i}{4}\sigma^{3}R_{12mi}y^{m}\,.
\end{align}
\end{subequations}
It is from the close identification of the spin connection resembling a gauge connection for a gauge covariant derivative, i.e., $\partial_a-\bar{\Omega}_a\rightarrow \partial_a+ieA_a$, that motivates the emergent fields, and in the vicinity of curvature in graphene the spin connection is physical~\cite{PhysRevD.92.125005}. We will examine two such cases with time dependence whereby components of the spin connection in $\bar{\Omega}_a$ give rise to curvature induced fields, one in the adiabatic case and the other in the nonadiabatic and high-frequency case. It turns out that they differ, with the high-frequency case representing a new class of emergent field. Before we analyze just how they emerge, let us spell out their decomposition from the above spin connections. Let us first look at the adiabatic case. What is essential here is the appearance of terms proportional to the spinor diagonal element, $i \sigma_3$. Let us go ahead and write their decomposition suggestively in the form of a pseudogauge field for spatial indices as
\begin{equation} \label{eq:our_pseudoA}
    \pazocal{A}_i^\text{Ad} \coloneqq  \frac{1}{4}R_{12mi}y^{m} \,.
\end{equation}
In two dimensions the Riemann tensor has only one degree of freedom and can be completely characterized by the Ricci scalar, and in (2+1) dimensions, the Riemann tensor has six degrees of freedom allowing for a Ricci decomposition,
\begin{equation}
    R_{\mu \nu \rho \sigma} = 2(g_{\mu [\rho}R_{\sigma]\nu}-g_{\nu [\rho}R_{\sigma]\mu}) - Rg_{\mu [\rho}g_{\sigma]\nu}\,.
\end{equation}
In the above decomposition let us write the above pseudogauge fields as
\begin{equation}
    \pazocal{A}_i^{\text{Ad}}= \frac{1}{8}(R\epsilon_{i m} + 4 R_{\hphantom{j}[i}^{j}\epsilon^{\vphantom{j}}_{m]j}) y^{m} \,.
\end{equation}
In the static two-dimensional limit, the Ricci tensor is completely determined by its scalar $R_{ij} = \tfrac{1}{2}Rg_{ij}$, hence only the vector potential $\pazocal{A}_{i}^{\text{Ad}} = \tfrac{1}{8}R\epsilon_{ik}y^k $ remains. This hints at the emergence of a uniform pseudo-magnetic field in symmetric gauge modulated solely by curvature. Indeed, under some considerations, the two-dimensional system embeds the dynamics of the Landau Hamiltonian~\cite{pseudoBgraphene}. 
The consideration of time dependency results, in addition to $\pazocal{A}_{i}^{\text{Ad}}$, into what could be a non-vanishing scalar potential $\pazocal{A}_{0}^{\text{Ad}}$ in $(1+2)$ dimensions. However, we will demonstrate in the next section that electric fields associated with the scalar potential are not present in the emergent system in the same way as they are for emergent magnetic fields. 

Next let us turn our attention to the high-frequency nonadiabatic case; we will find such emergent fields exist in Floquet materials, which we will discuss in Sec.~\ref{sec:Floquet}. To show their appearance let us write for the effective Hamiltonian in (2+1) dimensions,
\begin{equation}\label{eq:h_A}
    h_{\text{eff}} = \pazocal{H}_0 +i\sigma^{i}\bar{\Omega}_{i}+i\bar{\Omega}_{0} +h_{L}\,.
\end{equation}
Here the free Hamiltonian is $\pazocal{H}_0 = \sigma^ip_i$, and the remaining terms are represented by $h_L$. Note that we have also introduced the operator, $p_i=-i\partial_i$. We will eventually show the term $h_L$ introduces angular momentum dependent terms, with a specific example of FLRW space-time in Sec.~\ref{sec:FermiCoords}. The emergence of an entirely new pseudo-magnetic field comes from the identification that not only may $\bar{\Omega}_i$ contribute to a gauge-like coupling, but also $\bar{\Omega}_0$. Let us consider the first term in $\bar{\Omega}_0$, or $(1/2)\sigma^jR_{j00m}y^m$. Then note that one may write $\sigma^j=-i\epsilon^{ij}\sigma^3\sigma^i$ for $j=1,2$. In this way one can see that there is an additional gauge-like term that is proportional to $\sigma^3$ and that modifies Eq.~\eqref{eq:our_pseudoA}. For later usage, we go ahead and identify the combined term in a Floquet setting as
\begin{equation}\label{eq:Floqet_field_def}
    \pazocal{A}_i^\text{F}\coloneqq \frac{1}{4}R_{12mi}y^{m}+\epsilon_{ji}\frac{1}{2}R_{j00m}y^{m}\,,
\end{equation}
where notice the first term is given by Eq.~\eqref{eq:our_pseudoA}. Again let us perform a Ricci decomposition to find that
\begin{equation}
    \pazocal{A}_i^{\text{F}}= \frac{1}{2}\Bigl[\Bigl(R_{00}+\frac{3}{4}R\Bigr)\epsilon_{im}+R_{\hphantom{j}[i}^{j}\epsilon_{m]j}^{\vphantom{j}}+R_{m}^{\hphantom{m}j}\epsilon_{ji}\Bigr]y^{m}\,.
\end{equation}
Then we can see that the proposed field will be modified by time-dependent contributions directly with the addition of the $R_{00}$ and also the coefficient associated with the scalar curvature will be changed in contrast to Eq.~\eqref{eq:our_pseudoA}. While the effective Hermitian Hamiltonian given in Eq.~\eqref{eq:h_A} resembles the quantum mechanical equivalent, because of the spin dependent structure the relationship requires elucidation. An essential way to achieve this is by exploiting the quadratic form of the Dirac operator~\cite{pseudoBgraphene}, which has the added benefit of physical opacity, as we will show in the next section. Moreover in examining the quadratic form we may first analyze the adiabatic case given in Eq.~\eqref{eq:our_pseudoA}. Since this approach is conventional for the time-independent case, it will also provide insight into the coming sections relevant for Floquet environments.

Before introducing the quadratic form, let us remark on the evolution operator, describing fully the physics of our effective Hamiltonian system,
\begin{equation}\label{eq:unitary}
    U(y^0,0)=\pazocal{T}\exp\Bigl(-i \int^{y^0}_0dt'\, h_{\text{eff}} \Bigr)\,.
\end{equation}
To leading order in the curvature, one can determine that $[\bar{\Omega}_a,h_L]\simeq 0$, thus it is anticipated that the pseudogauge dynamics should be decoupled from those in $h_L$ according to the Baker–Campbell–Hausdorff formula. We will treat an expansion of $h_{\text{eff}}$ with rigor and phenomenologically in the context of Floquet engineering for graphene in Sec.~\ref{sec:Floquet}.

\section{Bispinor Green's Function}
\label{sec:QHam}

Above we examined an effective time-dependent flat space Hamiltonian visible by virtue of FNC from its curved space counterpart. One may equally well express a flat space effective theory stemming from a competing bispinor auxiliary fermionic Green's function that is quadratic in the Dirac operator suggested by deWitt (see, e.g., Refs.~\cite{dewitt1965dynamical,parker2009quantum}). The bispinor Green's function approach has proved important in the realization of time-independent pseudo-magnetic fields in strained graphene~\cite{pseudoBgraphene}; we will shortly demonstrate their emergence below. While we will employ the effective Hamiltonian derived in Eq.~\eqref{eq:h_A} for the purpose of Floquet analog realizations of high frequency time-dependent geometries, it is important that we draw the connection to the emergent fields in their conventionally understood quadratic form. 

To begin let us recall that $(\cgamma^\mu\nabla_\mu)^2=\Box -(1/4)R$, in accordance with the Lichnerowicz theorem~\cite{friedrich2000dirac}.
Application of which for massless fermions leads to the defining bispinor Green's function~\cite{parker2009quantum}, we write in FNC,
\begin{equation}\label{eq:Greens}
    \Bigl( \Box -\frac{1}{4}R 
    \Bigr)\,\mathcal{G}(y,y') = -\delta (y,y') \mathbb{1}_s\,.
\end{equation}
where $\delta(y,y'):=(-g)^{-1/2}\delta(y-y')$ is understood. To realize an effective quadratic flat space Green's function stemming from a time-dependent Hamiltonian we define the propagator $(-g)^{-1/4}\bar{\mathcal{G}}$. Such a redefinition is an important simplification in the construction of a momentum space bispinor Green's function~\cite{parker2009quantum}. The redefinition serves additionally as a means of achieving a Hermitian Hamiltonian in the static limit (from which the quadratic form may be found after squaring).
However, this feature for our time-dependent backgrounds is not necessarily the case; we will discover the quadratic form possesses anti-Hermitian terms. Last, as with the linear Hamiltonian examined earlier, usage of FNC in the bispinor Green's function is indispensable in the construction of its emergent form; this is demonstrated in RNC in~\cite{pseudoBgraphene} for the time-independent case. To introduce the new factors let us explicitly write the d'Alembertian as
\begin{align}\label{eq:dAl_def}
    \Box\mathcal{G} &=g^{\mu\nu}[\partial_{\mu , \nu}-(\partial_{\mu}\Omega_{\nu})-2\Omega_{(\mu}\partial_{\nu)} \nonumber \\
    &\hphantom{=g^{ij}}\;\, + \Omega_{\mu}\Omega_{\nu} - \Gamma_{\mu\nu}^{\lambda}(\partial_{\lambda}-\Omega_{\lambda})]\mathcal{G}\,.
\end{align}
Then, the redefined propagator can be shown to satisfy in FNC
\begin{equation}\label{eq:Greens2}
    \Bigl( \Box -\frac{1}{4}R +\bar{\Delta}
    \Bigr)\,\bar{\mathcal{G}}(y,0) = -\delta (y) \mathbb{1}_s\,,
\end{equation}
where we have conveniently chosen $y'=0$, and
\begin{align}
    \bar{\Delta}&=-\frac{1}{6}(\partial_{0}R_{lm}-2\partial_{0}R_{0l0m})y^{l}y^{m}\partial_{0}\nonumber\\
    &-\frac{1}{12}(\partial_{0}^{2}R_{lm}-2\partial_{0}^{2}R_{0l0m})y^{l}y^{m}\nonumber\\
    &+\frac{1}{3}\eta^{ij}(R_{im}-2R_{0i0m})y^{m}\partial_{j}+\frac{1}{6}(R-R_{00})\,.
\end{align}
The redefinition has augmented the quadratic operator with a new scalar addition.

To proceed let us make use of the FNC and explicitly express the operator up to linear order in curvature. Furthermore, we absorb $(-g^{00})^{-1}$ into the delta function. Then by application of Eq.~\eqref{eq:dAl_def} and by virtue of FNC, we find the bispinor Green's function and Eq.~\eqref{eq:Greens2} may be expressed as $\mathcal{D}_2\,\bar{\mathcal{G}}(y,0)=-\delta(y)\mathbb{1}_s$, where  
\begin{align}\label{eq:effGreens}
    \mathcal{D}_2&\coloneqq (\partial_{a}-\Omega_{a})^2 -\eta^{ab}\Gamma_{ab}^c\partial_{c}-\frac{1}{4}R +\bar{\Delta} \nonumber\\
    &+ y^{l}y^{m}\Bigl[R_{0l0m}\partial^{i}+\frac{1}{3}R_{\;l\;m}^{i\;j}\partial_{j}-\frac{4}{3}R_{0lim}\partial_{0}\Bigr]\partial_{i}\,
\end{align}
to linear order in the curvature (let us emphasize the $\Omega_a\Omega^a$ term should be disregarded). The spin and Christoffel connections are those given in Eqs.~\eqref{eq:SpinC_FNC} and~\eqref{eq:Cconn} respectively. The bispinor Green's function with Eq.~\eqref{eq:effGreens} has been expressed entirely in terms of flat space indices, and it represents the most general quadratic form for time-dependent emergent flat space systems from their curved counterparts, a hallmark of which is emergent pseudogauge fields. 

To begin, for reference, let us show the form of a bispinor Green's function in flat space with the adiabatic pseudogauge as written in Eq.~\eqref{eq:our_pseudoA}. Then in analogy to the Lichnerowicz theorem in flat space, one may write a quadratic in operator Green's function for massive fermions in arbitrary dimension, i.e., $[(\gamma_a\Pi^a)^2-m^2]\,\mathcal{G}_{\text{U}(1)}(y,0)=-\delta(y)$, where for U$(1)$ covariant derivative, $\Pi_a=\partial_a+ieA_a$ one has $(\gamma_a\Pi^a)^2= \Pi_a\Pi^a+(i/4)eF_{ab}\Sigma^{ab}$ for field strength $F_{ab}=\partial_a A_b-\partial_bA_a$. One can readily contrast this with an analogous expression for the spin connection in Eq.~\eqref{eq:effGreens}, found by virtue of the Lichnerowicz theorem and FNC written in flat space, $(\partial_{a}-\Omega_{a})^2-(1/4)R$. The spin connection possesses a similar form for the covariant operator as it does for the U$(1)$ case, albeit with spin structure, which in turn gives rise to the scalar curvature in place of $\Sigma_{ab}F^{ab}$. The key observation then is for the (2+1) dimensional case of the graphene sheet with $\Omega_a\rightarrow i\pazocal{A}_a^\text{Ad}\sigma_3$, as in Eq.~\eqref{eq:our_pseudoA}, one may more clearly recognize an emergent field structure than can be seen with a linear Dirac operator as demonstrated in Eq.~\eqref{eq:h_A}. This interpretation, however, comes with qualification. Let us decompose the would be emergent field curvature as
\begin{equation}
    \pazocal{F}_{ij}^{\text{Ad}}=\frac{1}{2}R_{12ij}\,,
\end{equation}
where $\pazocal{F}_{ab}^{\text{Ad}}=\partial_{a}\pazocal{A}_{b}^{\text{Ad}}-\partial_{b}\pazocal{A}_{a}^{\text{Ad}}$. Let us now consider the spin factor term using the above fields. One would expect to find $(i/4)\Sigma^{ab}\pazocal{F}^\text{Ad}_{ab}\sigma_3=-2\pazocal{F}_{12}^{\text{Ad}}$. One can immediately recognize in Eq.~\eqref{eq:effGreens}, that the corresponding field term is included by the way of the Ricci scalar and its augmentation from $\bar{\Delta}$. In passing let us comment on what would be an electric field identification; namely one would have  $\pazocal{F}_{0i}^{\text{Ad}}=-\frac{1}{4}(\partial_{0}R_{12im})y^{m}+\frac{1}{2}R_{120i}$. While the pseudogauge of such a term is present in the $(\partial_a-\Omega_a)^2$, no such corresponding field term is present in what would be the $(i/4)\Sigma^{0i}\pazocal{F}^\text{Ad}_{0i}\sigma_3$ term, therefore we cannot see pseudo-electric fields the same way one could see pseudo-magnetic fields. Let us also remark in passing that the $(R/12)$ term has also been attributed to an emergent and effective quadratic mass term~\cite{ChiralMassGap,InflationFromFerm}, rather than as part of an electromagnetic spin factor as we utilize here. The quadratic form of Eq.~\eqref{eq:effGreens}, while manifest in its connection to an emergent field structure, possesses many features that lie outside of the scope of this work such as non-Hermitian terms, but Eq.~\eqref{eq:effGreens} is amenable to an adiabatic analysis we will show with the concrete FLRW example in the coming section. For now, in its most general form let us, however, emphasize the emergence of homogeneous magnetic fields following the time-independent approach of~\cite{pseudoBgraphene}.

One may determine the time-independent limit of the quadratic Dirac operator, Eq.~\eqref{eq:effGreens}, as
\begin{equation}\label{eq:int_h}
    \mathcal{D}_2 \overset{\text{T.I.}}{\longrightarrow}-\partial_{0}^{2}+(\partial_{i}+i\sigma^{3}\pazocal{A}_{i}^{\text{Ad}})^{2}-\frac{1}{6}RL^{2}-\frac{1}{12}R\,,
\end{equation}
where we have introduced the angular momentum operator $L=y^{1}p^{2}-y^{2}p^{1}$. The appearance of $L^{2}$ corresponds to a frame dependent effect emerging at the space sector as a result of the choice of coordinates and is hence present for RNC as well as FNC. 

In this way, the $\mathcal{D}_2+\partial_0^2$ operator exactly coincides with the expression obtained by~\cite{pseudoBgraphene} for the effective quadratic Hamiltonian. 
Already looking at $\Sigma_{ab}$ terms in~\eqref{eq:int_h}, one can identify the appearance of a pseudo-magnetic field, $B_{\text{ps}}\coloneqq \hbar R/4e$, as reported in~\cite{pseudoBgraphene} and in~\cite{guinea2010energy} in the context of strain fields.

Finally, before introducing the FLRW metric, let us remark on the quadratic Hamiltonian. For the curved space Hamiltonian given in Eq.~\eqref{eq:Lin_Ham}, (not the effective Hamiltonian later determined), let us contrast $\pazocal{H}^2$ with the $\mathcal{D}_2$ operator. One would anticipate from the time-independent case that $\mathcal{D}_2 \subset \pazocal{H}^2$. To verify this, we draw our attention to 
\begin{align}
    \pazocal{H}^{2} &=  (g^{00})^{-1}\cgamma^i \cgamma^k \nabla_i \nabla_k + 2 \sigma^k g^{0i} \partial_{i, k} + \partial_{i}g_{00}\sigma^i \sigma^k \partial_{k} \nonumber \\
    &\hphantom{=}\, +i\{\pazocal{H}_0 , \Omega_0 \} + i\Omega_0 \pazocal{H}_0
    \label{eq:hamilt2}
\end{align}
This quadratic Hamiltonian fully contains the spatial sector of the quadratic Dirac operator,
\begin{align} \label{eq:relation_wD2}
    \frac{1}{g^{00}}\cgamma^{\mu}\nabla_{\mu} 
    \cgamma^{\nu}\nabla_{\nu} &= \partial_{0}^2 + \pazocal{H}^2 - \tfrac{1}{4}R_{0a}\sigma^a - \partial_{i}g_{00}\sigma^i \sigma^k \partial_k \nonumber \\
    &\hphantom{=}\, -i\pazocal{H}_0 [\Omega_0] -\partial_0 \Omega_0 \,.
\end{align}
One may directly contrast the above with $\mathcal{D}_2$, given in Eq.~\eqref{eq:effGreens}, with exception to the additive correction $\bar{\Delta}$ found from a rescaling of the Green's function by $(-g)^{-1/4}$. For static time-independent Hamiltonians Eq.~\eqref{eq:hamilt2} suffices to determine the dispersion relation; time dependence, on the other hand, spoils the equivalence between $\mathcal{D}_2+\partial_0^2$ and $\pazocal{H}^2$. Thus, for our adiabatic approach we directly consider Eq.~\eqref{eq:effGreens}.

\section{FLRW space-time in Fermi normal coordinates}
\label{sec:FermiCoords}

Above we demonstrated the emergence of a flat space-time Hermitian Hamiltonian from its curved space-time counterpart by virtue of FNC, which fully describes the time-dependent dynamics. In (2+1) dimensions it was furthermore shown in Eq.~\eqref{eq:our_pseudoA} and Eq.~\eqref{eq:Floqet_field_def}, that for small spatial inhomogeneities in the metric, components of the spin connection could be likened to emergent pseudogauge fields resembling electromagnetic fields. Let us more fully explore the time-dependent effective Hamiltonian with a concrete example of an FLRW metric in (2+1) dimensions. The metric is well describable in FNC, and moreover, models well time-dependent but small deformations of the membrane. 
In the context of the space-homogeneous FLRW geometry, the corrugations pertaining to this work then refer to time-dependent in-plane elastic stresses propagating throughout the membrane.

Let us characterize the FLRW metric in the absence of spatial curvature with conventional scalar factor, $a(x^0)$, such that
\begin{equation}
    \label{eq:FLRW_metric}
    ds^2 = -(dx^0)^2 + a(x^0)^2\delta_{ij}dx^i dx^j \,,
\end{equation}
with $i,j\in (1,2)$ in Cartesian coordinates. Let us remark at this point that we will later take $a(x^0)$ as an oscillatory profile, describing deformations about flat space-time that are periodic about $x^0$, for the purpose of experimental visibility in a Floquet environment; here the scalar factor is an arbitrary function. Our task now is to find a map between the conventional FLRW coordinates, which we denote with $x^\mu$, with those in FNC, which we have denoted as above with $y^\mu$.

To determine the map to FNC, let us follow the perturbative approach used in~\cite{Cooperstock:1998ny}, extending their range of validity to arbitrary scalar factor. First we require the geodesic equations that follow from Eq.~\eqref{eq:FLRW_metric}, these read
\begin{equation}
    \frac{d^{2}x^{0}}{d\tau^{2}}=-a\partial_{0}^{x}a\,\delta_{ij}\frac{dx^{i}}{d\tau}\frac{dx^{j}}{d\tau}\,,\quad\frac{d^{2}x^{i}}{d\tau^{2}}=-2\frac{\partial_{0}^{x}a}{a}\frac{dx^{0}}{d\tau}\frac{dx^{i}}{d\tau}\,,
\end{equation}
Here $\tau$ is the proper time. To differentiate with the FNC coordinates used elsewhere in this paper we denote partials with respect to conventional FLRW coordinates as $\partial^x_\mu=\partial/\partial x^\mu$. The above reduce to
\begin{equation}
    \frac{dx^0}{d\tau}=\sqrt{C_{2}+\frac{|C_{1}|^{2}}{a^{2}}}\,,\quad \frac{dx^{i}}{d\tau}=\frac{C_{1}^{i}}{a^{2}}\,,
\end{equation}
for constants $C_1^i$ and $C_2$ to be determined. A perturbative solution can be found to the above in $x^0(\tau)=x^0(0)+\tau dx^0/d\tau|_{\tau=0}+2!^{-1}\tau^2\,d^2x^0/d\tau^2|_{\tau=0}+3!^{-1}\tau^3\,d^3x^0/d\tau^3|_{\tau=0}$, where we truncate to $\pazocal{O}(\tau^4)$. A similar series with truncation is applied for $x^i(\tau)$. Then we take FNC at the beginning of the geodesic such that $y^0=x^0(0)$ and $y^i=a(y^0)(dx^i/d\tau)|_{\tau=0}$, where we have assumed for simplicity that $x^i(0)=0$. The initial conditions then set $C_1^i=a(y^0)y^i$ and $C_2=-|y^i|^2$. Last, one needs only to evaluate the perturbative solutions about a reference proper time, $\tau=1$, to determine the map between FNC and coordinates to $\pazocal{O}(y_i^4)$ as depicted in Eq.~\eqref{eq:FLRW_metric} as
\begin{equation}\label{eq:coordinate_map}
    x^{0}=y^{0}-\frac{|y^i|^{2}}{2}\frac{\dot{a}}{a}\,,\quad x^i=\frac{y^i}{a}\Bigl(1+\frac{|y^j|^{2}}{3}\frac{\dot{a}^{2}}{a^{2}}\Bigr)\,.
\end{equation}
Here and throughout ``dots" represents differentials with respect to FNC time, i.e., $\Dot{a}=\partial a/\partial y^0=\partial_0 a$. Using the above map one may determine the line element, Eq.~\eqref{eq:FLRW_metric}, expressed in FNC as
\begin{align}
    ds^{2}&=-\Bigl[1-|y^{i}|^{2}\frac{\ddot{a}}{a}\Bigr](dy^{0})^{2}+\frac{\dot{a}^{2}}{3a^{2}}(y^{i}dy^{i})^{2}\nonumber \\
    &\hphantom{=}\;+\Bigl[1-|y^{j}|^{2}\frac{\dot{a}^{2}}{3a^{2}}\Bigr] dy^{i}dy^{i}\,.
\end{align}

Equipped with the FNC metric one may readily confirm using Eq.~\eqref{eq:fnc_metric}, that to linear order in curvature one may determine the non-vanishing components of the Riemann tensor as
\begin{equation}
    R_{0i0i}=-(\Dot{H}+H^2)\,,\quad R_{1212}=H^2\,,
\end{equation}
where we have used the Hubble parameter, $H=\Dot{a}/a$. Likewise from which we can find the non-vanishing components of the Ricci tensor and scalar as
\begin{subequations}
\label{eq:Ricci_comp}
\begin{align}
    R_{00} &= - 2 (\Dot{H}+H^2)\,, \quad R_{ii} = \Dot{H}+ 2 H^2\,, \\
    R &= 4 \Dot{H}+ 6 H^2 \,,
\end{align}
\end{subequations}
whereas $R_{0 k}$ vanishes. The FLRW geometry in FNC is entirely characterizable with the Hubble constant and its derivatives.

With the above, let us first determine the effective Hamiltonian argued in Eq.~\eqref{eq:h_A}. Importantly we find for the augmented spin connections in Eq.~\eqref{eq:spinconn_FNCgen},
\begin{equation}\label{eq:aug_spin_FLRW}
    \bar{\Omega}_0=\frac{1}{2}(\Dot{H}+H^{2})\sigma^{i}y_{i}\,,\quad \bar{\Omega}_i=\frac{i}{4}H^{2}\sigma^{3}\epsilon_{ij}y^j\,,
\end{equation}
whose placement in the effective Hamiltonian may be expressed as
\begin{equation}\label{eq:flrw_h_1}
    i(\sigma^{i}\bar{\Omega}_{i}+\bar{\Omega}_{0})=\frac{i}{2}(\dot{H}+\tfrac{3}{2}H^{2})y^{i}\sigma_{i}\,.
\end{equation}
We will show in the following section the importance of such terms giving rise to an effective time-dependent gauge-like field in the high-frequency (nonadiabatic) regime analyzable with Floquet theory. In the linear effective Hamiltonian such a term appears to be imaginary, however it becomes real in the Floquet picture. Let us record, for later usage, the remaining term in the effective Hamiltonian as
\begin{equation}\label{eq:flrw_h_2}
    h_{L}=-\frac{1}{2}(\dot{H}+H^{2})y^{i}y_{i}\sigma^{j}p_{j}-\frac{1}{6}H^{2}y^{i}\sigma_{i}y^{j}p_{j}\,.
\end{equation}
Before discussing the nonadiabatic regime, let us utilize the FLRW metric to analyze the adiabatic case as argued applicable in the bispinor Green's function approach.

\subsection{Adiabatic pseudogauge Fields}

Let us first define adiabaticity as it is referenced here in the context of an FLRW geometry:
It is for Hubble parameters with weak FNC time dependence or rather
\begin{equation}\label{eq:adiabatic}
    \dot{H}\ll H\,.
\end{equation}
Let us emphasize that $H$ still may be time-dependent. In this approximation then one can determine that the $\mathcal{D}_2$ operator given in Eq.~\eqref{eq:effGreens} becomes
\begin{equation}\label{eq:D_2_ad}
    \mathcal{D}_{2}=\mathcal{D}_2^{\text{EM}}+\mathcal{D}_2^{\text{int}}\,,
\end{equation}
where we have for an electromagnetic like and an interaction-like term for the quadratic Dirac operator
\begin{align}
    \mathcal{D}_2^{\text{EM}}&=-\partial_{0}^{2}+(\partial_{i}+i\sigma^{3}\pazocal{A}_{i}^{\text{Ad}})^{2}-\frac{1}{12}R\,,\nonumber \\
    \mathcal{D}_2^{\text{int}}&=-H^{2}y^{i}\sigma_{i}\partial_{0}-H^{2}y^{i}y_{i}\partial^{j}\partial_{j}-\frac{1}{3}H^{2}L^{2}\,.
\end{align}
Recall that $L=y^{1}p^{2}-y^{2}p^{1}$. Notice the close similarity to the fully time-independent case given in Eq.~\eqref{eq:int_h}. The primary difference aside from a weak time dependence is that we have acquired a few additional terms. Let us emphasize that one cannot arrive at expression Eq.~\eqref{eq:int_h} in the time-independent limit of the Hubble parameter, since Eq.~\eqref{eq:int_h} is determined by omitting all time indices acting on the metric tensor.

To more clearly make the connection to its quantum mechanical counterpart let us express the Green's function with a DeWitt/Schwinger propertime integral~\cite{dewitt1965dynamical,PhysRev.82.664} as 
\begin{equation}
    \bar{\mathcal{G}}(y,0)=\lim_{\epsilon\rightarrow 0}\int^\infty_0 ds\,i\langle y|e^{-i[\mathcal{D}_2 +\epsilon(1-i)]s}|0\rangle\,.
\end{equation}
$\epsilon$ acts as a vanishing mass and provides propertime convergence in the IR limit. Then keeping terms only to the linear order in curvature we may use the Baker-Campbell-Hausdorff formula to write
\begin{equation}\label{eq:BCH_D_2}
    e^{-i\mathcal{D}_2s}\simeq e^{-i\mathcal{D}_2^{\text{EM}}s}e^{-i\mathcal{D}_2^{\text{int}}s}\,.
\end{equation}

Now let us write down the corresponding operator in (2+1) dimensional quantum electrodynamics
as
\begin{equation}
    e^{-i\slashed{\Pi}^2s}=e^{-i[-\partial_{0}^{2}+(\partial_{i}+ieA_{i})^{2}-\epsilon^{ij} eF_{ij}\sigma_3]s}\,.
\end{equation}
The essential difference to Eq.~\eqref{eq:BCH_D_2} is with the placement of $\sigma_3$ coupled to the gauge. Therefore with the help of projection operators,
\begin{equation}\label{eq:projection}
    P_\pm=(1/2)[1\pm \sigma^3]\,,
\end{equation}
one can determine that
\begin{equation}\label{eq:proj_G}
    P_\pm e^{-i\mathcal{D}_2^{\text{EM}}s}=P_\pm e^{-i[-\partial_{0}^{2}+(\partial_{i}\pm i\pazocal{A}_{i}^{\text{Ad}})^{2}-\frac{1}{12}R]s}\,,
\end{equation}
and equivalently for the electromagnetic case
\begin{equation}
    P_\pm e^{-i\slashed{\Pi}^2s}= P_\pm e^{-i[-\partial_{0}^{2}+(\partial_{i}+ieA_{i})^{2}\mp \epsilon^{ij} eF_{ij} ]s}\,.
\end{equation}
Then comparing the above two one can see that for a projection of $P_+$ of a curved space system leads to an equivalent electromagnetic system with positive coupling, $e$. Likewise for projection $P_-$, a system is represented with electromagnetic field, but in the opposite direction or simply taking, $e\rightarrow -e$. The drawback or rather, differences, between the two stems from the interaction-like terms in $\mathcal{D}_2^{\text{int}}$. Thus the pseudogauge fields of the curved space system can be realized with a projection of the Green's function as $P_\pm \bar{\mathcal{G}}(y,0)$ but differ from a purely electromagnetic setup with the accompaniment of an interaction-like term. To leading order, 
and weak time dependence in graphene, one may write $e^{-i\mathcal{D}_2^{\text{int}}s}\simeq 1-i\mathcal{D}_2^{\text{int}}s$. Rather than demonstrating emergent pseudogauge fields through curvature, in light of the above, if one were to demonstrate emergent curvature through real electromagnetic field, one could then see that such an interaction-like term may predict certain emergent curved space phenomena. We exploit this in the next section in a nonadiabatic setup through Floquet engineering. However, before introducing this let us briefly mention from a practical standpoint how one might achieve the adiabatic FLRW geometry in the lab.

Subtleties arise when trying to manufacture the FLRW geometry, Eq.~\eqref{eq:FLRW_metric}, in the laboratory to produce adiabatic fields. First, let us recall that intrinsic curvature is the curvature detectable to the observers within the manifold, which emerges from distortions of the metric. In contrast, extrinsic curvature corresponds to the curvature perceived as a result of how the $d$-dimensional surface is embedded in $\mathbb{R}^{d+1}$. In~\cite{PhysRevD.90.025006}, the issue of translating (3+1) dimensional (with speed of light $c$) Lorentz invariant laboratory coordinates to the embedded (2+1) dimensional (with Fermi velocity, $v_F$) graphene coordinates was exemplified for the cases of, e.g., Rindler and de Sitter space-times, wherein the shift to embedding coordinates implies the emergence of nonremovable singularities~\cite{AntiIorio}. In depth analyses, as conducted in~\cite{PhysRevD.90.025006}, in translating the embedded FLRW coordinates in Eq.~\eqref{eq:FLRW_metric}---we caution the FLRW (2+1) dimensional coordinates are again not the FNC---lies outside the scope of this work, however let us mention a few specific details concerning our setup. The FLRW geometry considered in this paper is conformally flat and has no Gaussian curvature and hence should not be subject to the issue of non-removable singularities when shifting to embedding coordinates. Furthermore, we have emphasized above the importance of using FNC to correctly describe our time-dependent Hamiltonian, and for applications to real graphene systems we remark that fortunately one may make use of the simpler of the FLRW geometry given in Eq.~\eqref{eq:FLRW_metric}, i.e., $x^\mu$ when considering the embedding. Once the FLRW coordinates have been established, to make the connection to FNC, i.e., $y^\mu$, we have supplied a map in Eq.~\eqref{eq:coordinate_map}; one must however determine $y^\mu(x^\nu)$ by inverting the expressions perturbatively in FNC. Thus the process from laboratory coordinates to the desired FNC goes as (3+1) laboratory $\to$ (2+1) $x^\mu \to$ (2+1) $y^\mu$. The delicate issue of time in the embedding, making it challenging for applications to, e.g., analog Hawking radiation in graphene, is skirted by the map, Eq.~\eqref{eq:coordinate_map} in that one may select a common laboratory coordinates and FLRW times.

\section{Floquet Induced High-Frequency Geometries}
\label{sec:Floquet}

We have shown in the adiabatic (or rather weakly modulated time in a FNC) picture that emergent gauge-like fields can be seen for the bispinor Green function. Specifically we considered an FLRW geometry leading to Eq.~\eqref{eq:D_2_ad}. We now turn our attention to the targeted high frequency and hence nonadiabatic regime, looking similarly at an FLRW geometry, which we study using Floquet engineering.

Floquet engineering is the manipulation of quantum systems through the use of driven external fields periodic in time; see~\cite{oka2019floquet} for a review. Floquet engineering has proved valuable in laser-driven systems~\cite{CHU20041}, strongly correlated systems~\cite{RevModPhys.86.779}, cold atoms in optical traps~\cite{RevModPhys.89.011004}, and even engineered to induce axial-vector fields~\cite{ebihara2016chiral}, to name a few.
In cosmology, the periodic time dependence of the inflaton or axion condensate backgrounds are known, via Floquet theory, to affect production rates in the form of parametric resonance~\cite{allahverdi2010reheating}. As argued above, usage of Floquet theory for the condensed matter realizations of high-frequency cosmological setups are intuitive. Not only may one interpret a pseudogauge field as arising from curvature induced geometries, but also one may interpret the reverse, namely artificial geometries as arising from induced electromagnetic gauge fields. Such approaches have also been studied for cold atoms in spatial geometries in~\cite{Boada_2011} and by Floquet analysis also in spatial geometries in~\cite{mishra2015floquet}; however, our FLRW geometry as well as temporal periodicity approaches are new and, moreover, give rise to a new class of emergent fields.

Let us specifically examine the previous FLRW metric in Eq.~\eqref{eq:FLRW_metric}, where we introduce temporal periodicity in the $a(x^0)$ factor. Let us also make use of the high-frequency valid linear effective Hermitian Hamiltonian with components given in Eqs.~\eqref{eq:flrw_h_1} and~\eqref{eq:flrw_h_2}. While we stress the importance of emergent space-time from induced periodic fields, it is conceptually simpler to proceed along the lines in previous sections beginning with curved space-time. Therefore, let us consider an observer in the laboratory frame who selects FLRW coordinates as given in Eq.~\eqref{eq:FLRW_metric}; periodicity then dictates that $h_\textrm{eff}(x^0)=h_\textrm{eff}(x^0+T)$. However, notice that the effective Hamiltonian has already been transformed to $\pazocal{O}(y_i^2)$ in FNC, therefore neglecting higher order contributions according to Eq.~\eqref{eq:coordinate_map}, one may also consider periodicity in FNC as
\begin{equation}
    h_\textrm{eff}(y^0)=h_\textrm{eff}(y^0+T)\,.
\end{equation}
Under a high-frequency Hamiltonian, an expansion of Eq.~\eqref{eq:unitary} is possible whereby one averages over the period giving way to a time-independent \textit{effective Floquet} Hamiltonian. Let us stress that the effective Hermitian Hamiltonian based on the scalar product in Eq.~\eqref{eq:h_A} is a different object. 

To perform the high-frequency expansion let us use a Magnus expansion~\cite{https://doi.org/10.1002/cpa.3160070404,*BLANES2009151,CoherentAver} for small $T$. Note that one may alternatively make use of the unitary equivalent van Vleck theory~\cite{PhysRev.33.467}. Here we use the approach as outlined in~\cite{CoherentAver}. Let us first write down the effective Floquet Hamiltonian as
\begin{equation}
    h_{\text{eff}}^{\text{F}}=\frac{i}{T}\textrm{ln}\, U(T,0)\,,
\end{equation}
where the unitary operator is given in Eq.~\eqref{eq:unitary}; we have also chosen an initial FNC time $y^0=0$ for convenience. A hallmark of the Floquet-Magnus formalism is that a mapping to a time-independent system is provided by averaging over the period; thus the eigenvalues of $h_{\textrm{eff}}^{\textrm{F}}$ represent Floquet pseudo-energies of a static system. Next, we gather Eq.~\eqref{eq:aug_spin_FLRW} and Eq.~\eqref{eq:flrw_h_2} into an interaction like term,
\begin{equation}
    h_\text{int}=\frac{i}{2}(\dot{H}+\tfrac{3}{2}H^{2})y^{i}\sigma_{i} +h_\text{L}\,,
\end{equation}
to which we can write the following unitary operator:
\begin{equation}
    U_{\text{int}}(y^{0},0)= \pazocal{T}e^{-i\int_{0}^{y^{0}}dy^{0}h_{\text{int}}}\,,
\end{equation}
that satisfies $U_{\text{int}}(T,0)=1$. Using the above we can write in the interaction picture, for $\tilde{h}_0(y^0)=U_{\text{int}}(y^{0},0)^\dagger \pazocal{H}_0U_{\text{int}}(y^{0},0)$,
\begin{equation}
    \widetilde{U}_0(y^0,0)=\pazocal{T}e^{-i\int_{0}^{y^{0}}dy^{0}\tilde{h}_{0}}
    \,.
\end{equation}
It has been shown that~\cite{Evans_1967} the original unitary operator of the effective Hamiltonian, Eq.~\eqref{eq:unitary}, follows from the product
\begin{equation}
    \widetilde{U}_0(y^0,0)U_{\text{int}}(y^{0},0)=U(y^{0},0)\,,
\end{equation}
and hence we can arrive at the effective Floquet Hamiltonian~\cite{CoherentAver}
\begin{equation}
    h_{\text{eff}}^{\text{F}}=\frac{i}{T}\textrm{ln}\, \widetilde{U}_0(T,0)\,.
\end{equation}
The above may be evaluated in a Magnus expansion, where we keep the leading order terms to $\pazocal{O}(T)$ in the Dyson series
\begin{align}
    h_{\text{eff}}^{\text{F}}&\simeq T^{-1}\int_{0}^{T}dy^{0}\tilde{h}_{0} \nonumber \\
    &\simeq \pazocal{H}_{0}+iT^{-1}\int_{0}^{T}dy^{0}\int_{0}^{y^{0}}dy'^{0}[h_{\text{int}},\pazocal{H}_{0}]\,.
\end{align}
The evaluation of the effective Floquet Hamiltonian has been reduced to that of a commutator between the interaction and free terms. Let us turn our attention to the commutator. Using the following identities:
\begin{subequations}
\begin{align}
    [y^i\sigma_i,\sigma^j p_j] &= 2i(1+L \sigma^3)\,, \\
    [y^iy_i\,\sigma^j  p_j,\sigma^k p_k] &= 2i\,\sigma^i y_i\,\sigma^j p_j\,, \\
    [y^i \sigma_i\,y^j p_j,\sigma^k p_k] &= i(2 L \sigma^3 + 3) \,y^i p_i + L \sigma^3 \,,
\end{align}
\end{subequations}
one can determine that the commutator becomes
\begin{equation}
    [h_{\text{int}},\pazocal{H}_{0}]=-\dot{H}-\frac{3}{2}H^{2}-i\Bigl(\dot{H}+\frac{3}{2}H^{2}\Bigr)y^{i}p_{i}+h_\text{L}^\text{C}
    \label{eq:commutator}\,.
\end{equation}
Here the factor
\begin{equation}\label{eq:h_L^C}
    h_\text{L}^\text{C}\coloneqq-\frac{1}{3}H^{2}\sigma^{3}[2L+iy^{i}Lp_{i}]
\end{equation}
arises from the commutator of the angular momentum dependent terms. A merit of the Floquet approach, in addition to providing a static formulation, is that the commutator is now diagonal, permitting some simplicities. Moreover, we may now make the connection to an emergent electromagnetic field.

The above form of the commutator shares close similarities with one under an electromagnetic gauge in a high frequency limit. To demonstrate this let us turn to the massless (2+1) dimensional Hamiltonian in a U$(1)$ electromagnetic field that reads $h^\text{A}=h^{\text{A}}_0+h_{\text{int}}^\text{A}$, where $h^\text{A}_{0}=v_{F}\sigma^{i}p_i=\pazocal{H}_0$ and $h_{\text{int}}^\text{A}=v_{F}e\sigma^{i}A_i$, such as exists in graphene for Fermi velocity, $v_F$. Note that we have assumed only a spatial gauge and that the electric potential is zero. The equivalent commutator stemming from a Magnus expansion for a high-frequency driven electric field is
\begin{equation}\label{eq:h_int_A}
    [h_{\text{int}}^\text{A},\pazocal{H}_{0}]=v_{F}^{2}e\{2i(A_{1}p_{2}-A_{2}p_{1})-F_{12}\}\sigma^{3}\,.
\end{equation}
Returning to the curved space expression, the commutator in Eq.~\eqref{eq:commutator} can be written in the suggestive form
\begin{equation}\label{eq:h_int_F}
    [h_{\text{int}},\pazocal{H}_{0}]=v_F^2\{2i(\pazocal{A}^{\text{F}}_1p_{2}-\pazocal{A}_{2}^\text{F} p_{1})-\pazocal{F}_{12}^\text{F}+h_\text{L}^\text{C}\}\,,
\end{equation}
where the emergent electromagnetic fields in the Floquet picture read
\begin{equation}
    \pazocal{A}_i^\text{F}=\frac{1}{2}\Bigl(\dot{H}+\frac{3}{2}H^{2}\Bigr)\epsilon_{ji}y^{j}\,,\quad \pazocal{F}_{12}^\text{F}=\dot{H}+\frac{3}{2}H^{2}\,.
\end{equation}
See also Eq.~\eqref{eq:Floqet_field_def}, as advertised before. We have also made visible the implicit Fermi velocity in graphene. By virtue of the Floquet picture we have identified a completely different induced electromagnetic field. In contrast to the adiabatic picture where we found Eq.~\eqref{eq:our_pseudoA}, we have as anticipated terms that depend on time derivatives. We also have a different coefficient in front of $H^2$, which is present due to an extra contribution coming from the $\bar{\Omega}_0$ factor (it is also interesting that here as well no emergent electric field is found from the $\bar{\Omega}_0$ factor). The key difference between the electromagnetic case in Eq.~\eqref{eq:h_int_A} and the curved space case in Eq.~\eqref{eq:h_int_F} is the presence of $\sigma^3$ coupled to the electromagnetic terms. We saw earlier with the adiabatic case in Eq.~\eqref{eq:proj_G}, that such an identification did in fact lead to emergent fields. Here too the same feature is present with $A_i\sigma^3\rightarrow \pazocal{A}^\text{F}_i$. In contrast with the adiabatic case, however, here we see that the electromagnetic case equivalent is proportional to $\sigma^3$ rather than the emergent curved space case that is proportional to the identity element. Since the emergent gauge fields are Abelian we find observables take a simple form.

With the identifications above, the effective Floquet Hamiltonian can be written as
\begin{align}
    h_\text{eff}^\text{F}&\simeq \mathcal{H}_0+[h_\pazocal{F}^\text{F}]_T+i[h_\text{L}^\text{C}]_T\,,\\
    h_\pazocal{F}^\text{F}&\coloneqq-2(\pazocal{A}^{\text{F}}_1p_{2}-\pazocal{A}_{2}^\text{F} p_{1})-i\pazocal{F}_{12}^\text{F}\,,
\end{align}
where $[f]_T\coloneqq (v_F^2/T)\int_{0}^{T}dy^{0}\int_{0}^{y^{0}}dy'^{0}f(y'^0)$ for arbitrary function $f$. Let us remark that while the above is not at first glance Hermitian, it is guaranteed to be so according to our Hermitian effective action, Eq.~\eqref{eq:h_A}, and the Magnus expansion. Using the effective Floquet Hamiltonian let us write for the arbitrary observable, $\pazocal{O}$, the expectation value with the projection operator given in Eq.~\eqref{eq:projection} as
\begin{equation}
    \langle\pazocal{O}P_\pm \rangle=\textrm{Tr}\,\pazocal{O}P_\pm e^{-iTh_{\text{eff}}^{\text{F}}}\,.
\end{equation}
Let us analyze the above in the high-frequency small $T$ limit, keeping only leading order terms upon factorizing according to the Baker–Campbell–Hausdorff formula. We find
\begin{equation}
    e^{-iTh_{\text{eff}}^{\text{F}}}\simeq e^{-iT[h_\pazocal{F}^\text{F}]_T}e^{-iT\mathcal{H}_0}e^{T[h_\text{L}^\text{C}]_T}\,,
\end{equation}
where we further assume $e^{T[h_\text{L}^\text{C}]_T}\simeq 1+T[h_\text{L}^\text{C}]_T$. The main observation here is that with the projection operator one may write
\begin{equation}
    P_\pm e^{-iT[h_\pazocal{F}^\text{F}]_T}=P_\pm e^{\mp iT[h_\pazocal{F}^\text{F}]_T\sigma^3}\,,
\end{equation}
and therefore connect to the equivalent electromagnetic interaction term with $\sigma^3$ coupling [as was similarly employed in the adiabatic case, Eq.~\eqref{eq:proj_G}]. Use of different projection operators leads to differing in sign electromagnetic terms. We may finally express our observable in curved space as
\begin{equation}\label{eq:observable_final}
    \langle\pazocal{O}P_\pm \rangle\simeq \textrm{Tr}\,(1+T[h_\text{L}^\text{C}]_T)\pazocal{O}P_\pm e^{-iT(\pazocal{H}_0\pm [h_\pazocal{F}^\text{F}]_T\sigma^3)}\,.
\end{equation}

Our goal is to now replicate the same observable, however only through the use of induced electromagnetic fields. We write down as before for arbitrary observable with electromagnetic field, $\pazocal{O}^\text{A}$, the following expectation value:
\begin{align}
    \langle \pazocal{O}^\text{A}P_\pm \rangle_\text{A} &= \textrm{Tr}\,\pazocal{O}^\text{A}P_\pm e^{-iT(\mathcal{H}_0+ e[h_\text{A}^\text{F}]_T\sigma^3)}\,,\label{eq:O_A}\\
    h_\text{A}^\text{F}&\coloneqq-2(A_1 p_{2}-A_{2} p_{1})-iF_{12}^\text{F}
\end{align}
with spin projection operator. Then it becomes clear by examining Eq.~\eqref{eq:h_L^C} that with the observable
\begin{equation}
    \pazocal{O}^\text{A}=(1\pm e T[h_L^\text{A}]_T)\pazocal{O}\,,
\end{equation}
where the angular momentum included operator is again 
\begin{equation}\label{h_L^A}
    h_\text{L}^\text{A}=-\frac{1}{3}H^{2}\sigma^{3}[2L+iy^{i}Lp_{i}]\,,
\end{equation}
the curved space observable in Eq.~\eqref{eq:observable_final} can be exactly reproduced using only high-frequency magnetic fields. One finds as anticipated that
\begin{equation}
    A_i=\pm e^{-1}\pazocal{A}^\text{F}_i\,,\quad F_{12}=\pm e^{-1}\pazocal{F}_{12}^\text{F}\,.
\end{equation}
More specifically, to arrive at the $P_+$ $(P_-)$ expression in Eq.~\eqref{eq:observable_final} one need only interpret the fields as going in the same (opposite) direction as those found from curved space. In practice, it is simplest to superimpose two high-frequency magnetic fields, one with field strength $\dot{H}$ and the other with $(3/2)H^2$, on to one another, and in this way the measurement of Eq.~\eqref{h_L^A} can be performed. Since one can furthermore write $\langle \pazocal{O}\rangle=\langle \pazocal{O}P_+\rangle+\langle \pazocal{O}P_-\rangle$, one can in theory make any measurement for the curved space analog system by performing two measurements of Eq.~\eqref{eq:O_A} for fields with opposing direction. The novelty and importance of the above identification is apparent in that one may simulate on a tabletop experiment an induced high-frequency space-time with known Floquet techniques.

\section{Conclusions}
\label{sec:conclusions}

We have studied time-dependent geometries with the (2+1) dimensional massless Dirac equation for target environment of graphene in both the adiabatic and nonadiabatic high-frequency cases. For both cases it was shown how, from a mapping to an effective quantum mechanical system, emergent pseudogauge fields may be recognized. For both approaches, FNC were used to supply one with a locally flat expansion good for making the connection to a quantum mechanical Hamiltonian, with intuitive understanding of observables in time-dependent geometries.
Also, for both approaches a generic FLRW metric was employed to make the connection to cosmological systems of interest.

It has been recently argued that spatial deformations of the lattice, such as topological defects or out-plane distortions, may result into higher order derivative terms leading to a vanishing spin connection which hinders the relationship between a tight-binding model and the curved Dirac equation~(see \cite{PhysRevB.105.195412} and~\cite{PhysRevD.92.125005,PhysRevB.106.157401,*PhysRevB.106.157402}). 
However, the intrinsic curvature of the FLRW geometry we use in our work depends solely on a continuous time variable and therefore is not subject to such corrections. The corrugations in our setup are due to a simple elastic deformation. Let us also remark that in our work the pseudogauge fields and spin connection are dual to one another; however in~\cite{PhysRevB.105.195412}, an Abelian strain gauge field enters quite differently and separate to a non-Abelian torsion-free spin connection. It would, however, be interesting to explore geometries beyond the FLRW which encompass additional spatial deformations.

For the adiabatic case, we used the bispinor Green's function approach, which is known to predict homogeneous pseudo-magnetic fields in a time-independent geometry. We extended our understanding of such emergent pseudogauge fields to encompass small variations in FNC time for an FLRW metric such that $\dot{H}\ll H$, for Hubble parameter $H$. It was shown that the emergent temporally inhomogeneous pseudo-magnetic fields agree with their static counterparts.

For the nonadiabatic and high-frequency case we made use of the Floquet theorem for periodically driven systems. We determined an entirely new class of pseudogauge field existing at high-frequency, which differs from the adiabatic one through contributions coming from the temporal part of the Hermitic corrected spin connection, $\bar{\Omega}_0$. It is thought the electromagnetic field, conversely, may be produced in the laboratory onto graphene, to give rise to an emergent curved space setting thought valuable to study on a tabletop experiment. Some potential areas of interest include rotating neutron stars, compact binaries, early universe quantum fluctuations, and, in particular stochastic gravitational wave background (SGWB) during the inflationary period whose gravitational wave bath encompasses a frequency span of at least 20 orders of magnitude.

\begin{acknowledgments}
We would like to thank James Quach, Kazuya Mameda, Toby Wiseman and Matthew M. Roberts for valuable discussions.
\end{acknowledgments}

\bibliography{references}

\begin{thebibliography}{67}%
\makeatletter
\providecommand \@ifxundefined [1]{%
 \@ifx{#1\undefined}
}%
\providecommand \@ifnum [1]{%
 \ifnum #1\expandafter \@firstoftwo
 \else \expandafter \@secondoftwo
 \fi
}%
\providecommand \@ifx [1]{%
 \ifx #1\expandafter \@firstoftwo
 \else \expandafter \@secondoftwo
 \fi
}%
\providecommand \natexlab [1]{#1}%
\providecommand \enquote  [1]{``#1''}%
\providecommand \bibnamefont  [1]{#1}%
\providecommand \bibfnamefont [1]{#1}%
\providecommand \citenamefont [1]{#1}%
\providecommand \href@noop [0]{\@secondoftwo}%
\providecommand \href [0]{\begingroup \@sanitize@url \@href}%
\providecommand \@href[1]{\@@startlink{#1}\@@href}%
\providecommand \@@href[1]{\endgroup#1\@@endlink}%
\providecommand \@sanitize@url [0]{\catcode `\\12\catcode `\$12\catcode
  `\&12\catcode `\#12\catcode `\^12\catcode `\_12\catcode `\%12\relax}%
\providecommand \@@startlink[1]{}%
\providecommand \@@endlink[0]{}%
\providecommand \url  [0]{\begingroup\@sanitize@url \@url }%
\providecommand \@url [1]{\endgroup\@href {#1}{\urlprefix }}%
\providecommand \urlprefix  [0]{URL }%
\providecommand \Eprint [0]{\href }%
\providecommand \doibase [0]{http://dx.doi.org/}%
\providecommand \selectlanguage [0]{\@gobble}%
\providecommand \bibinfo  [0]{\@secondoftwo}%
\providecommand \bibfield  [0]{\@secondoftwo}%
\providecommand \translation [1]{[#1]}%
\providecommand \BibitemOpen [0]{}%
\providecommand \bibitemStop [0]{}%
\providecommand \bibitemNoStop [0]{.\EOS\space}%
\providecommand \EOS [0]{\spacefactor3000\relax}%
\providecommand \BibitemShut  [1]{\csname bibitem#1\endcsname}%
\let\auto@bib@innerbib\@empty
\bibitem [{\citenamefont {Novoselov}\ \emph {et~al.}(2005)\citenamefont
  {Novoselov}, \citenamefont {Geim}, \citenamefont {Morozov}, \citenamefont
  {Jiang}, \citenamefont {Katsnelson}, \citenamefont {Grigorieva},
  \citenamefont {Dubonos},\ and\ \citenamefont {Firsov}}]{Novoselov2005}%
  \BibitemOpen
  \bibfield  {author} {\bibinfo {author} {\bibfnamefont {K.~S.}\ \bibnamefont
  {Novoselov}}, \bibinfo {author} {\bibfnamefont {A.~K.}\ \bibnamefont {Geim}},
  \bibinfo {author} {\bibfnamefont {S.~V.}\ \bibnamefont {Morozov}}, \bibinfo
  {author} {\bibfnamefont {D.}~\bibnamefont {Jiang}}, \bibinfo {author}
  {\bibfnamefont {M.~I.}\ \bibnamefont {Katsnelson}}, \bibinfo {author}
  {\bibfnamefont {I.~V.}\ \bibnamefont {Grigorieva}}, \bibinfo {author}
  {\bibfnamefont {S.~V.}\ \bibnamefont {Dubonos}}, \ and\ \bibinfo {author}
  {\bibfnamefont {A.~A.}\ \bibnamefont {Firsov}},\ }\href {\doibase
  10.1038/nature04233} {\bibfield  {journal} {\bibinfo  {journal} {Nature}\
  }\textbf {\bibinfo {volume} {438}},\ \bibinfo {pages} {197} (\bibinfo {year}
  {2005})}\BibitemShut {NoStop}%
\bibitem [{\citenamefont {Geim}\ and\ \citenamefont
  {Novoselov}(2007)}]{Geim2007}%
  \BibitemOpen
  \bibfield  {author} {\bibinfo {author} {\bibfnamefont {A.~K.}\ \bibnamefont
  {Geim}}\ and\ \bibinfo {author} {\bibfnamefont {K.~S.}\ \bibnamefont
  {Novoselov}},\ }\href {\doibase 10.1038/nmat1849} {\bibfield  {journal}
  {\bibinfo  {journal} {Nature Materials}\ }\textbf {\bibinfo {volume} {6}},\
  \bibinfo {pages} {183} (\bibinfo {year} {2007})}\BibitemShut {NoStop}%
\bibitem [{\citenamefont {Iorio}\ and\ \citenamefont
  {Lambiase}(2012)}]{IORIO2012334}%
  \BibitemOpen
  \bibfield  {author} {\bibinfo {author} {\bibfnamefont {A.}~\bibnamefont
  {Iorio}}\ and\ \bibinfo {author} {\bibfnamefont {G.}~\bibnamefont
  {Lambiase}},\ }\href {\doibase
  https://doi.org/10.1016/j.physletb.2012.08.023} {\bibfield  {journal}
  {\bibinfo  {journal} {Physics Letters B}\ }\textbf {\bibinfo {volume}
  {716}},\ \bibinfo {pages} {334} (\bibinfo {year} {2012})}\BibitemShut
  {NoStop}%
\bibitem [{\citenamefont {Iorio}\ and\ \citenamefont
  {Lambiase}(2014)}]{PhysRevD.90.025006}%
  \BibitemOpen
  \bibfield  {author} {\bibinfo {author} {\bibfnamefont {A.}~\bibnamefont
  {Iorio}}\ and\ \bibinfo {author} {\bibfnamefont {G.}~\bibnamefont
  {Lambiase}},\ }\href {\doibase 10.1103/PhysRevD.90.025006} {\bibfield
  {journal} {\bibinfo  {journal} {Phys. Rev. D}\ }\textbf {\bibinfo {volume}
  {90}},\ \bibinfo {pages} {025006} (\bibinfo {year} {2014})}\BibitemShut
  {NoStop}%
\bibitem [{\citenamefont {Cvetič}\ and\ \citenamefont
  {Gibbons}(2012)}]{CVETIC20122617}%
  \BibitemOpen
  \bibfield  {author} {\bibinfo {author} {\bibfnamefont {M.}~\bibnamefont
  {Cvetič}}\ and\ \bibinfo {author} {\bibfnamefont {G.}~\bibnamefont
  {Gibbons}},\ }\href {\doibase https://doi.org/10.1016/j.aop.2012.05.013}
  {\bibfield  {journal} {\bibinfo  {journal} {Annals of Physics}\ }\textbf
  {\bibinfo {volume} {327}},\ \bibinfo {pages} {2617} (\bibinfo {year}
  {2012})}\BibitemShut {NoStop}%
\bibitem [{\citenamefont {González}\ and\ \citenamefont
  {Herrero}(2010)}]{GONZALEZ2010426}%
  \BibitemOpen
  \bibfield  {author} {\bibinfo {author} {\bibfnamefont {J.}~\bibnamefont
  {González}}\ and\ \bibinfo {author} {\bibfnamefont {J.}~\bibnamefont
  {Herrero}},\ }\href {\doibase
  https://doi.org/10.1016/j.nuclphysb.2009.09.028} {\bibfield  {journal}
  {\bibinfo  {journal} {Nuclear Physics B}\ }\textbf {\bibinfo {volume}
  {825}},\ \bibinfo {pages} {426} (\bibinfo {year} {2010})}\BibitemShut
  {NoStop}%
\bibitem [{\citenamefont {Morozov}\ \emph {et~al.}(2006)\citenamefont
  {Morozov}, \citenamefont {Novoselov}, \citenamefont {Katsnelson},
  \citenamefont {Schedin}, \citenamefont {Ponomarenko}, \citenamefont {Jiang},\
  and\ \citenamefont {Geim}}]{PhysRevLett.97.016801}%
  \BibitemOpen
  \bibfield  {author} {\bibinfo {author} {\bibfnamefont {S.~V.}\ \bibnamefont
  {Morozov}}, \bibinfo {author} {\bibfnamefont {K.~S.}\ \bibnamefont
  {Novoselov}}, \bibinfo {author} {\bibfnamefont {M.~I.}\ \bibnamefont
  {Katsnelson}}, \bibinfo {author} {\bibfnamefont {F.}~\bibnamefont {Schedin}},
  \bibinfo {author} {\bibfnamefont {L.~A.}\ \bibnamefont {Ponomarenko}},
  \bibinfo {author} {\bibfnamefont {D.}~\bibnamefont {Jiang}}, \ and\ \bibinfo
  {author} {\bibfnamefont {A.~K.}\ \bibnamefont {Geim}},\ }\href {\doibase
  10.1103/PhysRevLett.97.016801} {\bibfield  {journal} {\bibinfo  {journal}
  {Phys. Rev. Lett.}\ }\textbf {\bibinfo {volume} {97}},\ \bibinfo {pages}
  {016801} (\bibinfo {year} {2006})}\BibitemShut {NoStop}%
\bibitem [{\citenamefont {Morpurgo}\ and\ \citenamefont
  {Guinea}(2006)}]{PhysRevLett.97.196804}%
  \BibitemOpen
  \bibfield  {author} {\bibinfo {author} {\bibfnamefont {A.~F.}\ \bibnamefont
  {Morpurgo}}\ and\ \bibinfo {author} {\bibfnamefont {F.}~\bibnamefont
  {Guinea}},\ }\href {\doibase 10.1103/PhysRevLett.97.196804} {\bibfield
  {journal} {\bibinfo  {journal} {Phys. Rev. Lett.}\ }\textbf {\bibinfo
  {volume} {97}},\ \bibinfo {pages} {196804} (\bibinfo {year}
  {2006})}\BibitemShut {NoStop}%
\bibitem [{\citenamefont {Levy}\ \emph {et~al.}(2010)\citenamefont {Levy},
  \citenamefont {Burke}, \citenamefont {Meaker}, \citenamefont {Panlasigui},
  \citenamefont {Zettl}, \citenamefont {Guinea}, \citenamefont {Neto},\ and\
  \citenamefont {Crommie}}]{doi:10.1126/science.1191700}%
  \BibitemOpen
  \bibfield  {author} {\bibinfo {author} {\bibfnamefont {N.}~\bibnamefont
  {Levy}}, \bibinfo {author} {\bibfnamefont {S.~A.}\ \bibnamefont {Burke}},
  \bibinfo {author} {\bibfnamefont {K.~L.}\ \bibnamefont {Meaker}}, \bibinfo
  {author} {\bibfnamefont {M.}~\bibnamefont {Panlasigui}}, \bibinfo {author}
  {\bibfnamefont {A.}~\bibnamefont {Zettl}}, \bibinfo {author} {\bibfnamefont
  {F.}~\bibnamefont {Guinea}}, \bibinfo {author} {\bibfnamefont {A.~H.~C.}\
  \bibnamefont {Neto}}, \ and\ \bibinfo {author} {\bibfnamefont {M.~F.}\
  \bibnamefont {Crommie}},\ }\href {\doibase 10.1126/science.1191700}
  {\bibfield  {journal} {\bibinfo  {journal} {Science}\ }\textbf {\bibinfo
  {volume} {329}},\ \bibinfo {pages} {544} (\bibinfo {year} {2010})},\ \Eprint
  {http://arxiv.org/abs/https://www.science.org/doi/pdf/10.1126/science.1191700}
  {https://www.science.org/doi/pdf/10.1126/science.1191700} \BibitemShut
  {NoStop}%
\bibitem [{\citenamefont {de~Juan}\ \emph {et~al.}(2007)\citenamefont
  {de~Juan}, \citenamefont {Cortijo},\ and\ \citenamefont
  {Vozmediano}}]{PhysRevB.76.165409}%
  \BibitemOpen
  \bibfield  {author} {\bibinfo {author} {\bibfnamefont {F.}~\bibnamefont
  {de~Juan}}, \bibinfo {author} {\bibfnamefont {A.}~\bibnamefont {Cortijo}}, \
  and\ \bibinfo {author} {\bibfnamefont {M.~A.~H.}\ \bibnamefont
  {Vozmediano}},\ }\href {\doibase 10.1103/PhysRevB.76.165409} {\bibfield
  {journal} {\bibinfo  {journal} {Phys. Rev. B}\ }\textbf {\bibinfo {volume}
  {76}},\ \bibinfo {pages} {165409} (\bibinfo {year} {2007})}\BibitemShut
  {NoStop}%
\bibitem [{\citenamefont {Vozmediano}\ \emph {et~al.}(2008)\citenamefont
  {Vozmediano}, \citenamefont {de~Juan},\ and\ \citenamefont
  {Cortijo}}]{Vozmediano_2008}%
  \BibitemOpen
  \bibfield  {author} {\bibinfo {author} {\bibfnamefont {M.~A.~H.}\
  \bibnamefont {Vozmediano}}, \bibinfo {author} {\bibfnamefont
  {F.}~\bibnamefont {de~Juan}}, \ and\ \bibinfo {author} {\bibfnamefont
  {A.}~\bibnamefont {Cortijo}},\ }\href {\doibase
  10.1088/1742-6596/129/1/012001} {\bibfield  {journal} {\bibinfo  {journal}
  {Journal of Physics: Conference Series}\ }\textbf {\bibinfo {volume} {129}},\
  \bibinfo {pages} {012001} (\bibinfo {year} {2008})}\BibitemShut {NoStop}%
\bibitem [{\citenamefont {{de Juan}}\ \emph {et~al.}(2010)\citenamefont {{de
  Juan}}, \citenamefont {Cortijo},\ and\ \citenamefont
  {Vozmediano}}]{DEJUAN2010625}%
  \BibitemOpen
  \bibfield  {author} {\bibinfo {author} {\bibfnamefont {F.}~\bibnamefont {{de
  Juan}}}, \bibinfo {author} {\bibfnamefont {A.}~\bibnamefont {Cortijo}}, \
  and\ \bibinfo {author} {\bibfnamefont {M.~A.}\ \bibnamefont {Vozmediano}},\
  }\href {\doibase https://doi.org/10.1016/j.nuclphysb.2009.11.012} {\bibfield
  {journal} {\bibinfo  {journal} {Nuclear Physics B}\ }\textbf {\bibinfo
  {volume} {828}},\ \bibinfo {pages} {625} (\bibinfo {year}
  {2010})}\BibitemShut {NoStop}%
\bibitem [{\citenamefont {de~Juan}\ \emph {et~al.}(2013)\citenamefont
  {de~Juan}, \citenamefont {Ma\~nes},\ and\ \citenamefont
  {Vozmediano}}]{PhysRevB.87.165131}%
  \BibitemOpen
  \bibfield  {author} {\bibinfo {author} {\bibfnamefont {F.}~\bibnamefont
  {de~Juan}}, \bibinfo {author} {\bibfnamefont {J.~L.}\ \bibnamefont
  {Ma\~nes}}, \ and\ \bibinfo {author} {\bibfnamefont {M.~A.~H.}\ \bibnamefont
  {Vozmediano}},\ }\href {\doibase 10.1103/PhysRevB.87.165131} {\bibfield
  {journal} {\bibinfo  {journal} {Phys. Rev. B}\ }\textbf {\bibinfo {volume}
  {87}},\ \bibinfo {pages} {165131} (\bibinfo {year} {2013})}\BibitemShut
  {NoStop}%
\bibitem [{\citenamefont {Gonz\'alez}\ \emph {et~al.}(1992)\citenamefont
  {Gonz\'alez}, \citenamefont {Guinea},\ and\ \citenamefont
  {Vozmediano}}]{PhysRevLett.69.172}%
  \BibitemOpen
  \bibfield  {author} {\bibinfo {author} {\bibfnamefont {J.}~\bibnamefont
  {Gonz\'alez}}, \bibinfo {author} {\bibfnamefont {F.}~\bibnamefont {Guinea}},
  \ and\ \bibinfo {author} {\bibfnamefont {M.~A.~H.}\ \bibnamefont
  {Vozmediano}},\ }\href {\doibase 10.1103/PhysRevLett.69.172} {\bibfield
  {journal} {\bibinfo  {journal} {Phys. Rev. Lett.}\ }\textbf {\bibinfo
  {volume} {69}},\ \bibinfo {pages} {172} (\bibinfo {year} {1992})}\BibitemShut
  {NoStop}%
\bibitem [{\citenamefont {Gonzalez}\ \emph {et~al.}(1993)\citenamefont
  {Gonzalez}, \citenamefont {Guinea},\ and\ \citenamefont
  {Vozmediano}}]{gonzalez1993electronic}%
  \BibitemOpen
  \bibfield  {author} {\bibinfo {author} {\bibfnamefont {J.}~\bibnamefont
  {Gonzalez}}, \bibinfo {author} {\bibfnamefont {F.}~\bibnamefont {Guinea}}, \
  and\ \bibinfo {author} {\bibfnamefont {M.~A.}\ \bibnamefont {Vozmediano}},\
  }\href@noop {} {\bibfield  {journal} {\bibinfo  {journal} {Nuclear Physics
  B}\ }\textbf {\bibinfo {volume} {406}},\ \bibinfo {pages} {771} (\bibinfo
  {year} {1993})}\BibitemShut {NoStop}%
\bibitem [{\citenamefont {de~Juan}(2013)}]{PhysRevB.87.125419}%
  \BibitemOpen
  \bibfield  {author} {\bibinfo {author} {\bibfnamefont {F.}~\bibnamefont
  {de~Juan}},\ }\href {\doibase 10.1103/PhysRevB.87.125419} {\bibfield
  {journal} {\bibinfo  {journal} {Phys. Rev. B}\ }\textbf {\bibinfo {volume}
  {87}},\ \bibinfo {pages} {125419} (\bibinfo {year} {2013})}\BibitemShut
  {NoStop}%
\bibitem [{\citenamefont {Cortijo}\ \emph {et~al.}(2015)\citenamefont
  {Cortijo}, \citenamefont {Ferreir\'os}, \citenamefont {Landsteiner},\ and\
  \citenamefont {Vozmediano}}]{PhysRevLett.115.177202}%
  \BibitemOpen
  \bibfield  {author} {\bibinfo {author} {\bibfnamefont {A.}~\bibnamefont
  {Cortijo}}, \bibinfo {author} {\bibfnamefont {Y.}~\bibnamefont
  {Ferreir\'os}}, \bibinfo {author} {\bibfnamefont {K.}~\bibnamefont
  {Landsteiner}}, \ and\ \bibinfo {author} {\bibfnamefont {M.~A.~H.}\
  \bibnamefont {Vozmediano}},\ }\href {\doibase 10.1103/PhysRevLett.115.177202}
  {\bibfield  {journal} {\bibinfo  {journal} {Phys. Rev. Lett.}\ }\textbf
  {\bibinfo {volume} {115}},\ \bibinfo {pages} {177202} (\bibinfo {year}
  {2015})}\BibitemShut {NoStop}%
\bibitem [{\citenamefont {Soto-Garrido}\ and\ \citenamefont
  {Muñoz}(2018)}]{Soto-Garrido_2018}%
  \BibitemOpen
  \bibfield  {author} {\bibinfo {author} {\bibfnamefont {R.}~\bibnamefont
  {Soto-Garrido}}\ and\ \bibinfo {author} {\bibfnamefont {E.}~\bibnamefont
  {Muñoz}},\ }\href {\doibase 10.1088/1361-648X/aaba07} {\bibfield  {journal}
  {\bibinfo  {journal} {Journal of Physics: Condensed Matter}\ }\textbf
  {\bibinfo {volume} {30}},\ \bibinfo {pages} {195302} (\bibinfo {year}
  {2018})}\BibitemShut {NoStop}%
\bibitem [{\citenamefont {Soto-Garrido}\ \emph {et~al.}(2020)\citenamefont
  {Soto-Garrido}, \citenamefont {Mu\~noz},\ and\ \citenamefont {Juri\ifmmode
  \check{c}\else \v{c}\fi{}i\ifmmode~\acute{c}\else
  \'{c}\fi{}}}]{PhysRevResearch.2.012043}%
  \BibitemOpen
  \bibfield  {author} {\bibinfo {author} {\bibfnamefont {R.}~\bibnamefont
  {Soto-Garrido}}, \bibinfo {author} {\bibfnamefont {E.}~\bibnamefont
  {Mu\~noz}}, \ and\ \bibinfo {author} {\bibfnamefont {V.}~\bibnamefont
  {Juri\ifmmode \check{c}\else \v{c}\fi{}i\ifmmode~\acute{c}\else
  \'{c}\fi{}}},\ }\href {\doibase 10.1103/PhysRevResearch.2.012043} {\bibfield
  {journal} {\bibinfo  {journal} {Phys. Rev. Res.}\ }\textbf {\bibinfo {volume}
  {2}},\ \bibinfo {pages} {012043} (\bibinfo {year} {2020})}\BibitemShut
  {NoStop}%
\bibitem [{\citenamefont {Bonilla}\ and\ \citenamefont
  {Muñoz}(2022)}]{nano12203711}%
  \BibitemOpen
  \bibfield  {author} {\bibinfo {author} {\bibfnamefont {D.}~\bibnamefont
  {Bonilla}}\ and\ \bibinfo {author} {\bibfnamefont {E.}~\bibnamefont
  {Muñoz}},\ }\href {\doibase 10.3390/nano12203711} {\bibfield  {journal}
  {\bibinfo  {journal} {Nanomaterials}\ }\textbf {\bibinfo {volume} {12}}
  (\bibinfo {year} {2022}),\ 10.3390/nano12203711}\BibitemShut {NoStop}%
\bibitem [{\citenamefont {Vozmediano}\ \emph {et~al.}(2010)\citenamefont
  {Vozmediano}, \citenamefont {Katsnelson},\ and\ \citenamefont
  {Guinea}}]{VOZMEDIANO2010109}%
  \BibitemOpen
  \bibfield  {author} {\bibinfo {author} {\bibfnamefont {M.}~\bibnamefont
  {Vozmediano}}, \bibinfo {author} {\bibfnamefont {M.}~\bibnamefont
  {Katsnelson}}, \ and\ \bibinfo {author} {\bibfnamefont {F.}~\bibnamefont
  {Guinea}},\ }\href {\doibase https://doi.org/10.1016/j.physrep.2010.07.003}
  {\bibfield  {journal} {\bibinfo  {journal} {Physics Reports}\ }\textbf
  {\bibinfo {volume} {496}},\ \bibinfo {pages} {109} (\bibinfo {year}
  {2010})}\BibitemShut {NoStop}%
\bibitem [{\citenamefont {Pe\~na}\ and\ \citenamefont
  {Mu\~noz}(2015)}]{PhysRevE.91.052152}%
  \BibitemOpen
  \bibfield  {author} {\bibinfo {author} {\bibfnamefont {F.~J.}\ \bibnamefont
  {Pe\~na}}\ and\ \bibinfo {author} {\bibfnamefont {E.}~\bibnamefont
  {Mu\~noz}},\ }\href {\doibase 10.1103/PhysRevE.91.052152} {\bibfield
  {journal} {\bibinfo  {journal} {Phys. Rev. E}\ }\textbf {\bibinfo {volume}
  {91}},\ \bibinfo {pages} {052152} (\bibinfo {year} {2015})}\BibitemShut
  {NoStop}%
\bibitem [{\citenamefont {Castro-Villarreal}\ and\ \citenamefont
  {Ruiz-S\'anchez}(2017)}]{pseudoBgraphene}%
  \BibitemOpen
  \bibfield  {author} {\bibinfo {author} {\bibfnamefont {P.}~\bibnamefont
  {Castro-Villarreal}}\ and\ \bibinfo {author} {\bibfnamefont {R.}~\bibnamefont
  {Ruiz-S\'anchez}},\ }\href {\doibase 10.1103/PhysRevB.95.125432} {\bibfield
  {journal} {\bibinfo  {journal} {Phys. Rev. B}\ }\textbf {\bibinfo {volume}
  {95}},\ \bibinfo {pages} {125432} (\bibinfo {year} {2017})}\BibitemShut
  {NoStop}%
\bibitem [{\citenamefont {Oka}\ and\ \citenamefont
  {Kitamura}(2019)}]{oka2019floquet}%
  \BibitemOpen
  \bibfield  {author} {\bibinfo {author} {\bibfnamefont {T.}~\bibnamefont
  {Oka}}\ and\ \bibinfo {author} {\bibfnamefont {S.}~\bibnamefont {Kitamura}},\
  }\href {https://doi.org/10.1146/annurev-conmatphys-031218-013423} {\bibfield
  {journal} {\bibinfo  {journal} {Annual Review of Condensed Matter Physics}\
  }\textbf {\bibinfo {volume} {10}},\ \bibinfo {pages} {387} (\bibinfo {year}
  {2019})}\BibitemShut {NoStop}%
\bibitem [{\citenamefont {Nicolis}\ \emph {et~al.}(2009)\citenamefont
  {Nicolis}, \citenamefont {Rattazzi},\ and\ \citenamefont
  {Trincherini}}]{PhysRevD.79.064036}%
  \BibitemOpen
  \bibfield  {author} {\bibinfo {author} {\bibfnamefont {A.}~\bibnamefont
  {Nicolis}}, \bibinfo {author} {\bibfnamefont {R.}~\bibnamefont {Rattazzi}}, \
  and\ \bibinfo {author} {\bibfnamefont {E.}~\bibnamefont {Trincherini}},\
  }\href {\doibase 10.1103/PhysRevD.79.064036} {\bibfield  {journal} {\bibinfo
  {journal} {Phys. Rev. D}\ }\textbf {\bibinfo {volume} {79}},\ \bibinfo
  {pages} {064036} (\bibinfo {year} {2009})}\BibitemShut {NoStop}%
\bibitem [{\citenamefont {Allahverdi}\ \emph {et~al.}(2010)\citenamefont
  {Allahverdi}, \citenamefont {Brandenberger}, \citenamefont {Cyr-Racine},\
  and\ \citenamefont {Mazumdar}}]{allahverdi2010reheating}%
  \BibitemOpen
  \bibfield  {author} {\bibinfo {author} {\bibfnamefont {R.}~\bibnamefont
  {Allahverdi}}, \bibinfo {author} {\bibfnamefont {R.}~\bibnamefont
  {Brandenberger}}, \bibinfo {author} {\bibfnamefont {F.-Y.}\ \bibnamefont
  {Cyr-Racine}}, \ and\ \bibinfo {author} {\bibfnamefont {A.}~\bibnamefont
  {Mazumdar}},\ }\href
  {https://www.annualreviews.org/doi/full/10.1146/annurev.nucl.012809.104511}
  {\bibfield  {journal} {\bibinfo  {journal} {Annual Review of Nuclear and
  Particle Science}\ }\textbf {\bibinfo {volume} {60}},\ \bibinfo {pages} {27}
  (\bibinfo {year} {2010})}\BibitemShut {NoStop}%
\bibitem [{\citenamefont {Benisty}\ \emph {et~al.}(2019)\citenamefont
  {Benisty}, \citenamefont {Guendelman}, \citenamefont {Saridakis},
  \citenamefont {Stoecker}, \citenamefont {Struckmeier},\ and\ \citenamefont
  {Vasak}}]{InflationFromFerm}%
  \BibitemOpen
  \bibfield  {author} {\bibinfo {author} {\bibfnamefont {D.}~\bibnamefont
  {Benisty}}, \bibinfo {author} {\bibfnamefont {E.~I.}\ \bibnamefont
  {Guendelman}}, \bibinfo {author} {\bibfnamefont {E.~N.}\ \bibnamefont
  {Saridakis}}, \bibinfo {author} {\bibfnamefont {H.}~\bibnamefont {Stoecker}},
  \bibinfo {author} {\bibfnamefont {J.}~\bibnamefont {Struckmeier}}, \ and\
  \bibinfo {author} {\bibfnamefont {D.}~\bibnamefont {Vasak}},\ }\href
  {\doibase 10.1103/PhysRevD.100.043523} {\bibfield  {journal} {\bibinfo
  {journal} {Phys. Rev. D}\ }\textbf {\bibinfo {volume} {100}},\ \bibinfo
  {pages} {043523} (\bibinfo {year} {2019})}\BibitemShut {NoStop}%
\bibitem [{\citenamefont {Adshead}\ and\ \citenamefont
  {Sfakianakis}(2016)}]{AxionInflation}%
  \BibitemOpen
  \bibfield  {author} {\bibinfo {author} {\bibfnamefont {P.}~\bibnamefont
  {Adshead}}\ and\ \bibinfo {author} {\bibfnamefont {E.~I.}\ \bibnamefont
  {Sfakianakis}},\ }\href {\doibase 10.1103/PhysRevLett.116.091301} {\bibfield
  {journal} {\bibinfo  {journal} {Phys. Rev. Lett.}\ }\textbf {\bibinfo
  {volume} {116}},\ \bibinfo {pages} {091301} (\bibinfo {year}
  {2016})}\BibitemShut {NoStop}%
\bibitem [{\citenamefont {Maleknejad}(2020)}]{maleknejad2020dark}%
  \BibitemOpen
  \bibfield  {author} {\bibinfo {author} {\bibfnamefont {A.}~\bibnamefont
  {Maleknejad}},\ }\href@noop {} {\bibfield  {journal} {\bibinfo  {journal}
  {Journal of High Energy Physics}\ }\textbf {\bibinfo {volume} {2020}},\
  \bibinfo {pages} {1} (\bibinfo {year} {2020})}\BibitemShut {NoStop}%
\bibitem [{\citenamefont {Parker}(1980{\natexlab{a}})}]{ParkerLinearH}%
  \BibitemOpen
  \bibfield  {author} {\bibinfo {author} {\bibfnamefont {L.}~\bibnamefont
  {Parker}},\ }\href {\doibase 10.1103/PhysRevD.22.1922} {\bibfield  {journal}
  {\bibinfo  {journal} {Phys. Rev. D}\ }\textbf {\bibinfo {volume} {22}},\
  \bibinfo {pages} {1922} (\bibinfo {year} {1980}{\natexlab{a}})}\BibitemShut
  {NoStop}%
\bibitem [{\citenamefont {Parker}(1980{\natexlab{b}})}]{ParkerPRL}%
  \BibitemOpen
  \bibfield  {author} {\bibinfo {author} {\bibfnamefont {L.}~\bibnamefont
  {Parker}},\ }\href {\doibase 10.1103/PhysRevLett.44.1559} {\bibfield
  {journal} {\bibinfo  {journal} {Phys. Rev. Lett.}\ }\textbf {\bibinfo
  {volume} {44}},\ \bibinfo {pages} {1559} (\bibinfo {year}
  {1980}{\natexlab{b}})}\BibitemShut {NoStop}%
\bibitem [{\citenamefont {Fermi}(1921)}]{fermi1921sull}%
  \BibitemOpen
  \bibfield  {author} {\bibinfo {author} {\bibfnamefont {E.}~\bibnamefont
  {Fermi}},\ }\href@noop {} {\bibfield  {journal} {\bibinfo  {journal} {Il
  Nuovo Cimento (1911-1923)}\ }\textbf {\bibinfo {volume} {22}},\ \bibinfo
  {pages} {176} (\bibinfo {year} {1921})}\BibitemShut {NoStop}%
\bibitem [{\citenamefont {Marzlin}(1994)}]{Marzlin:1994wc}%
  \BibitemOpen
  \bibfield  {author} {\bibinfo {author} {\bibfnamefont {K.-P.}\ \bibnamefont
  {Marzlin}},\ }\href {\doibase 10.1007/BF02108003} {\bibfield  {journal}
  {\bibinfo  {journal} {Gen. Rel. Grav.}\ }\textbf {\bibinfo {volume} {26}},\
  \bibinfo {pages} {619} (\bibinfo {year} {1994})},\ \Eprint
  {http://arxiv.org/abs/gr-qc/9402010} {arXiv:gr-qc/9402010} \BibitemShut
  {NoStop}%
\bibitem [{\citenamefont {Nesterov}(1999)}]{Nesterov_1999}%
  \BibitemOpen
  \bibfield  {author} {\bibinfo {author} {\bibfnamefont {A.~I.}\ \bibnamefont
  {Nesterov}},\ }\href {\doibase 10.1088/0264-9381/16/2/011} {\bibfield
  {journal} {\bibinfo  {journal} {Classical and Quantum Gravity}\ }\textbf
  {\bibinfo {volume} {16}},\ \bibinfo {pages} {465} (\bibinfo {year}
  {1999})}\BibitemShut {NoStop}%
\bibitem [{\citenamefont {Cao}\ \emph {et~al.}(2020)\citenamefont {Cao},
  \citenamefont {Feng}, \citenamefont {Han}, \citenamefont {Gao}, \citenamefont
  {Hue~Ly}, \citenamefont {Xu},\ and\ \citenamefont {Lu}}]{cao2020elastic}%
  \BibitemOpen
  \bibfield  {author} {\bibinfo {author} {\bibfnamefont {K.}~\bibnamefont
  {Cao}}, \bibinfo {author} {\bibfnamefont {S.}~\bibnamefont {Feng}}, \bibinfo
  {author} {\bibfnamefont {Y.}~\bibnamefont {Han}}, \bibinfo {author}
  {\bibfnamefont {L.}~\bibnamefont {Gao}}, \bibinfo {author} {\bibfnamefont
  {T.}~\bibnamefont {Hue~Ly}}, \bibinfo {author} {\bibfnamefont
  {Z.}~\bibnamefont {Xu}}, \ and\ \bibinfo {author} {\bibfnamefont
  {Y.}~\bibnamefont {Lu}},\ }\href@noop {} {\bibfield  {journal} {\bibinfo
  {journal} {Nature communications}\ }\textbf {\bibinfo {volume} {11}},\
  \bibinfo {pages} {284} (\bibinfo {year} {2020})}\BibitemShut {NoStop}%
\bibitem [{\citenamefont {Lee}\ \emph {et~al.}(2008)\citenamefont {Lee},
  \citenamefont {Wei}, \citenamefont {Kysar},\ and\ \citenamefont
  {Hone}}]{lee2008measurement}%
  \BibitemOpen
  \bibfield  {author} {\bibinfo {author} {\bibfnamefont {C.}~\bibnamefont
  {Lee}}, \bibinfo {author} {\bibfnamefont {X.}~\bibnamefont {Wei}}, \bibinfo
  {author} {\bibfnamefont {J.~W.}\ \bibnamefont {Kysar}}, \ and\ \bibinfo
  {author} {\bibfnamefont {J.}~\bibnamefont {Hone}},\ }\href@noop {} {\bibfield
   {journal} {\bibinfo  {journal} {science}\ }\textbf {\bibinfo {volume}
  {321}},\ \bibinfo {pages} {385} (\bibinfo {year} {2008})}\BibitemShut
  {NoStop}%
\bibitem [{\citenamefont {Zhang}\ \emph {et~al.}(2005)\citenamefont {Zhang},
  \citenamefont {Tan}, \citenamefont {Stormer},\ and\ \citenamefont
  {Kim}}]{Zhang2005}%
  \BibitemOpen
  \bibfield  {author} {\bibinfo {author} {\bibfnamefont {Y.}~\bibnamefont
  {Zhang}}, \bibinfo {author} {\bibfnamefont {Y.-W.}\ \bibnamefont {Tan}},
  \bibinfo {author} {\bibfnamefont {H.~L.}\ \bibnamefont {Stormer}}, \ and\
  \bibinfo {author} {\bibfnamefont {P.}~\bibnamefont {Kim}},\ }\href {\doibase
  10.1038/nature04235} {\bibfield  {journal} {\bibinfo  {journal} {Nature}\
  }\textbf {\bibinfo {volume} {438}},\ \bibinfo {pages} {201} (\bibinfo {year}
  {2005})}\BibitemShut {NoStop}%
\bibitem [{\citenamefont {Yan}\ \emph {et~al.}(2013)\citenamefont {Yan},
  \citenamefont {Chu}, \citenamefont {Yan}, \citenamefont {Liu}, \citenamefont
  {Meng}, \citenamefont {Yang}, \citenamefont {Fan}, \citenamefont {Wang},
  \citenamefont {Dou}, \citenamefont {Zhang}, \citenamefont {Liu},
  \citenamefont {Nie},\ and\ \citenamefont {He}}]{PhysRevB.87.075405}%
  \BibitemOpen
  \bibfield  {author} {\bibinfo {author} {\bibfnamefont {H.}~\bibnamefont
  {Yan}}, \bibinfo {author} {\bibfnamefont {Z.-D.}\ \bibnamefont {Chu}},
  \bibinfo {author} {\bibfnamefont {W.}~\bibnamefont {Yan}}, \bibinfo {author}
  {\bibfnamefont {M.}~\bibnamefont {Liu}}, \bibinfo {author} {\bibfnamefont
  {L.}~\bibnamefont {Meng}}, \bibinfo {author} {\bibfnamefont {M.}~\bibnamefont
  {Yang}}, \bibinfo {author} {\bibfnamefont {Y.}~\bibnamefont {Fan}}, \bibinfo
  {author} {\bibfnamefont {J.}~\bibnamefont {Wang}}, \bibinfo {author}
  {\bibfnamefont {R.-F.}\ \bibnamefont {Dou}}, \bibinfo {author} {\bibfnamefont
  {Y.}~\bibnamefont {Zhang}}, \bibinfo {author} {\bibfnamefont
  {Z.}~\bibnamefont {Liu}}, \bibinfo {author} {\bibfnamefont {J.-C.}\
  \bibnamefont {Nie}}, \ and\ \bibinfo {author} {\bibfnamefont
  {L.}~\bibnamefont {He}},\ }\href {\doibase 10.1103/PhysRevB.87.075405}
  {\bibfield  {journal} {\bibinfo  {journal} {Phys. Rev. B}\ }\textbf {\bibinfo
  {volume} {87}},\ \bibinfo {pages} {075405} (\bibinfo {year}
  {2013})}\BibitemShut {NoStop}%
\bibitem [{\citenamefont {Katsnelson}\ \emph {et~al.}(2006)\citenamefont
  {Katsnelson}, \citenamefont {Novoselov},\ and\ \citenamefont
  {Geim}}]{Katsnelson2006}%
  \BibitemOpen
  \bibfield  {author} {\bibinfo {author} {\bibfnamefont {M.~I.}\ \bibnamefont
  {Katsnelson}}, \bibinfo {author} {\bibfnamefont {K.~S.}\ \bibnamefont
  {Novoselov}}, \ and\ \bibinfo {author} {\bibfnamefont {A.~K.}\ \bibnamefont
  {Geim}},\ }\href {\doibase 10.1038/nphys384} {\bibfield  {journal} {\bibinfo
  {journal} {Nature Physics}\ }\textbf {\bibinfo {volume} {2}},\ \bibinfo
  {pages} {620} (\bibinfo {year} {2006})}\BibitemShut {NoStop}%
\bibitem [{Note1()}]{Note1}%
  \BibitemOpen
  \bibinfo {note} {Let us point out that if one were to model intrinsic
  (membrane) properties rather than the extrinsic (embedding) in graphene then
  it has been advocated to use both Dirac points, rather the single one assumed
  here~\cite {OLIVALEYVA20152645,AntiIorio}}\BibitemShut {NoStop}%
\bibitem [{\citenamefont {Manasse}\ and\ \citenamefont
  {Misner}(1963)}]{manasse1963fermi}%
  \BibitemOpen
  \bibfield  {author} {\bibinfo {author} {\bibfnamefont {F.}~\bibnamefont
  {Manasse}}\ and\ \bibinfo {author} {\bibfnamefont {C.~W.}\ \bibnamefont
  {Misner}},\ }\href@noop {} {\bibfield  {journal} {\bibinfo  {journal}
  {Journal of mathematical physics}\ }\textbf {\bibinfo {volume} {4}},\
  \bibinfo {pages} {735} (\bibinfo {year} {1963})}\BibitemShut {NoStop}%
\bibitem [{\citenamefont {Chicone}\ and\ \citenamefont
  {Mashhoon}(2006)}]{dSFermiCoords}%
  \BibitemOpen
  \bibfield  {author} {\bibinfo {author} {\bibfnamefont {C.}~\bibnamefont
  {Chicone}}\ and\ \bibinfo {author} {\bibfnamefont {B.}~\bibnamefont
  {Mashhoon}},\ }\href {\doibase 10.1103/PhysRevD.74.064019} {\bibfield
  {journal} {\bibinfo  {journal} {Phys. Rev. D}\ }\textbf {\bibinfo {volume}
  {74}},\ \bibinfo {pages} {064019} (\bibinfo {year} {2006})}\BibitemShut
  {NoStop}%
\bibitem [{\citenamefont {Huang}\ and\ \citenamefont
  {Parker}(2009)}]{PhysRevD.79.024020}%
  \BibitemOpen
  \bibfield  {author} {\bibinfo {author} {\bibfnamefont {X.}~\bibnamefont
  {Huang}}\ and\ \bibinfo {author} {\bibfnamefont {L.}~\bibnamefont {Parker}},\
  }\href {\doibase 10.1103/PhysRevD.79.024020} {\bibfield  {journal} {\bibinfo
  {journal} {Phys. Rev. D}\ }\textbf {\bibinfo {volume} {79}},\ \bibinfo
  {pages} {024020} (\bibinfo {year} {2009})}\BibitemShut {NoStop}%
\bibitem [{\citenamefont {Iorio}\ and\ \citenamefont
  {Pais}(2015)}]{PhysRevD.92.125005}%
  \BibitemOpen
  \bibfield  {author} {\bibinfo {author} {\bibfnamefont {A.}~\bibnamefont
  {Iorio}}\ and\ \bibinfo {author} {\bibfnamefont {P.}~\bibnamefont {Pais}},\
  }\href {\doibase 10.1103/PhysRevD.92.125005} {\bibfield  {journal} {\bibinfo
  {journal} {Phys. Rev. D}\ }\textbf {\bibinfo {volume} {92}},\ \bibinfo
  {pages} {125005} (\bibinfo {year} {2015})}\BibitemShut {NoStop}%
\bibitem [{\citenamefont {DeWitt}(1965)}]{dewitt1965dynamical}%
  \BibitemOpen
  \bibfield  {author} {\bibinfo {author} {\bibfnamefont {B.~S.}\ \bibnamefont
  {DeWitt}},\ }\href@noop {} {\emph {\bibinfo {title} {Dynamical theory of
  groups and fields}}}\ (\bibinfo  {publisher} {Gordon and Breach},\ \bibinfo
  {year} {1965})\BibitemShut {NoStop}%
\bibitem [{\citenamefont {Parker}\ and\ \citenamefont
  {Toms}(2009)}]{parker2009quantum}%
  \BibitemOpen
  \bibfield  {author} {\bibinfo {author} {\bibfnamefont {L.}~\bibnamefont
  {Parker}}\ and\ \bibinfo {author} {\bibfnamefont {D.}~\bibnamefont {Toms}},\
  }\href@noop {} {\emph {\bibinfo {title} {Quantum field theory in curved
  spacetime: quantized fields and gravity}}}\ (\bibinfo  {publisher} {Cambridge
  university press},\ \bibinfo {year} {2009})\BibitemShut {NoStop}%
\bibitem [{\citenamefont {Friedrich}(2000)}]{friedrich2000dirac}%
  \BibitemOpen
  \bibfield  {author} {\bibinfo {author} {\bibfnamefont {T.}~\bibnamefont
  {Friedrich}},\ }\href@noop {} {\emph {\bibinfo {title} {Dirac operators in
  Riemannian geometry}}},\ Vol.~\bibinfo {volume} {25}\ (\bibinfo  {publisher}
  {American Mathematical Soc.},\ \bibinfo {year} {2000})\BibitemShut {NoStop}%
\bibitem [{\citenamefont {Flachi}\ and\ \citenamefont
  {Fukushima}(2014)}]{ChiralMassGap}%
  \BibitemOpen
  \bibfield  {author} {\bibinfo {author} {\bibfnamefont {A.}~\bibnamefont
  {Flachi}}\ and\ \bibinfo {author} {\bibfnamefont {K.}~\bibnamefont
  {Fukushima}},\ }\href {\doibase 10.1103/PhysRevLett.113.091102} {\bibfield
  {journal} {\bibinfo  {journal} {Phys. Rev. Lett.}\ }\textbf {\bibinfo
  {volume} {113}},\ \bibinfo {pages} {091102} (\bibinfo {year}
  {2014})}\BibitemShut {NoStop}%
\bibitem [{\citenamefont {Guinea}\ \emph {et~al.}(2010)\citenamefont {Guinea},
  \citenamefont {Katsnelson},\ and\ \citenamefont {Geim}}]{guinea2010energy}%
  \BibitemOpen
  \bibfield  {author} {\bibinfo {author} {\bibfnamefont {F.}~\bibnamefont
  {Guinea}}, \bibinfo {author} {\bibfnamefont {M.}~\bibnamefont {Katsnelson}},
  \ and\ \bibinfo {author} {\bibfnamefont {A.}~\bibnamefont {Geim}},\
  }\href@noop {} {\bibfield  {journal} {\bibinfo  {journal} {Nature Physics}\
  }\textbf {\bibinfo {volume} {6}},\ \bibinfo {pages} {30} (\bibinfo {year}
  {2010})}\BibitemShut {NoStop}%
\bibitem [{\citenamefont {Cooperstock}\ \emph {et~al.}(1998)\citenamefont
  {Cooperstock}, \citenamefont {Faraoni},\ and\ \citenamefont
  {Vollick}}]{Cooperstock:1998ny}%
  \BibitemOpen
  \bibfield  {author} {\bibinfo {author} {\bibfnamefont {F.~I.}\ \bibnamefont
  {Cooperstock}}, \bibinfo {author} {\bibfnamefont {V.}~\bibnamefont
  {Faraoni}}, \ and\ \bibinfo {author} {\bibfnamefont {D.~N.}\ \bibnamefont
  {Vollick}},\ }\href {\doibase 10.1086/305956} {\bibfield  {journal} {\bibinfo
   {journal} {The Astrophysical Journal}\ }\textbf {\bibinfo {volume} {503}},\
  \bibinfo {pages} {61} (\bibinfo {year} {1998})}\BibitemShut {NoStop}%
\bibitem [{\citenamefont {Schwinger}(1951)}]{PhysRev.82.664}%
  \BibitemOpen
  \bibfield  {author} {\bibinfo {author} {\bibfnamefont {J.}~\bibnamefont
  {Schwinger}},\ }\href {\doibase 10.1103/PhysRev.82.664} {\bibfield  {journal}
  {\bibinfo  {journal} {Phys. Rev.}\ }\textbf {\bibinfo {volume} {82}},\
  \bibinfo {pages} {664} (\bibinfo {year} {1951})}\BibitemShut {NoStop}%
\bibitem [{\citenamefont {Iorio}\ and\ \citenamefont {Pais}(2018)}]{AntiIorio}%
  \BibitemOpen
  \bibfield  {author} {\bibinfo {author} {\bibfnamefont {A.}~\bibnamefont
  {Iorio}}\ and\ \bibinfo {author} {\bibfnamefont {P.}~\bibnamefont {Pais}},\
  }\href {\doibase https://doi.org/10.1016/j.aop.2018.09.011} {\bibfield
  {journal} {\bibinfo  {journal} {Annals of Physics}\ }\textbf {\bibinfo
  {volume} {398}},\ \bibinfo {pages} {265} (\bibinfo {year}
  {2018})}\BibitemShut {NoStop}%
\bibitem [{\citenamefont {Chu}\ and\ \citenamefont {Telnov}(2004)}]{CHU20041}%
  \BibitemOpen
  \bibfield  {author} {\bibinfo {author} {\bibfnamefont {S.-I.}\ \bibnamefont
  {Chu}}\ and\ \bibinfo {author} {\bibfnamefont {D.~A.}\ \bibnamefont
  {Telnov}},\ }\href {\doibase https://doi.org/10.1016/j.physrep.2003.10.001}
  {\bibfield  {journal} {\bibinfo  {journal} {Physics Reports}\ }\textbf
  {\bibinfo {volume} {390}},\ \bibinfo {pages} {1} (\bibinfo {year}
  {2004})}\BibitemShut {NoStop}%
\bibitem [{\citenamefont {Aoki}\ \emph {et~al.}(2014)\citenamefont {Aoki},
  \citenamefont {Tsuji}, \citenamefont {Eckstein}, \citenamefont {Kollar},
  \citenamefont {Oka},\ and\ \citenamefont {Werner}}]{RevModPhys.86.779}%
  \BibitemOpen
  \bibfield  {author} {\bibinfo {author} {\bibfnamefont {H.}~\bibnamefont
  {Aoki}}, \bibinfo {author} {\bibfnamefont {N.}~\bibnamefont {Tsuji}},
  \bibinfo {author} {\bibfnamefont {M.}~\bibnamefont {Eckstein}}, \bibinfo
  {author} {\bibfnamefont {M.}~\bibnamefont {Kollar}}, \bibinfo {author}
  {\bibfnamefont {T.}~\bibnamefont {Oka}}, \ and\ \bibinfo {author}
  {\bibfnamefont {P.}~\bibnamefont {Werner}},\ }\href {\doibase
  10.1103/RevModPhys.86.779} {\bibfield  {journal} {\bibinfo  {journal} {Rev.
  Mod. Phys.}\ }\textbf {\bibinfo {volume} {86}},\ \bibinfo {pages} {779}
  (\bibinfo {year} {2014})}\BibitemShut {NoStop}%
\bibitem [{\citenamefont {Eckardt}(2017)}]{RevModPhys.89.011004}%
  \BibitemOpen
  \bibfield  {author} {\bibinfo {author} {\bibfnamefont {A.}~\bibnamefont
  {Eckardt}},\ }\href {\doibase 10.1103/RevModPhys.89.011004} {\bibfield
  {journal} {\bibinfo  {journal} {Rev. Mod. Phys.}\ }\textbf {\bibinfo {volume}
  {89}},\ \bibinfo {pages} {011004} (\bibinfo {year} {2017})}\BibitemShut
  {NoStop}%
\bibitem [{\citenamefont {Ebihara}\ \emph {et~al.}(2016)\citenamefont
  {Ebihara}, \citenamefont {Fukushima},\ and\ \citenamefont
  {Oka}}]{ebihara2016chiral}%
  \BibitemOpen
  \bibfield  {author} {\bibinfo {author} {\bibfnamefont {S.}~\bibnamefont
  {Ebihara}}, \bibinfo {author} {\bibfnamefont {K.}~\bibnamefont {Fukushima}},
  \ and\ \bibinfo {author} {\bibfnamefont {T.}~\bibnamefont {Oka}},\ }\href
  {\doibase 10.1103/PhysRevB.93.155107} {\bibfield  {journal} {\bibinfo
  {journal} {Phys. Rev. B}\ }\textbf {\bibinfo {volume} {93}},\ \bibinfo
  {pages} {155107} (\bibinfo {year} {2016})}\BibitemShut {NoStop}%
\bibitem [{\citenamefont {Boada}\ \emph {et~al.}(2011)\citenamefont {Boada},
  \citenamefont {Celi}, \citenamefont {Latorre},\ and\ \citenamefont
  {Lewenstein}}]{Boada_2011}%
  \BibitemOpen
  \bibfield  {author} {\bibinfo {author} {\bibfnamefont {O.}~\bibnamefont
  {Boada}}, \bibinfo {author} {\bibfnamefont {A.}~\bibnamefont {Celi}},
  \bibinfo {author} {\bibfnamefont {J.~I.}\ \bibnamefont {Latorre}}, \ and\
  \bibinfo {author} {\bibfnamefont {M.}~\bibnamefont {Lewenstein}},\ }\href
  {\doibase 10.1088/1367-2630/13/3/035002} {\bibfield  {journal} {\bibinfo
  {journal} {New Journal of Physics}\ }\textbf {\bibinfo {volume} {13}},\
  \bibinfo {pages} {035002} (\bibinfo {year} {2011})}\BibitemShut {NoStop}%
\bibitem [{\citenamefont {Mishra}\ \emph {et~al.}(2015)\citenamefont {Mishra},
  \citenamefont {Sarkar},\ and\ \citenamefont
  {Bandyopadhyay}}]{mishra2015floquet}%
  \BibitemOpen
  \bibfield  {author} {\bibinfo {author} {\bibfnamefont {T.}~\bibnamefont
  {Mishra}}, \bibinfo {author} {\bibfnamefont {T.~G.}\ \bibnamefont {Sarkar}},
  \ and\ \bibinfo {author} {\bibfnamefont {J.~N.}\ \bibnamefont
  {Bandyopadhyay}},\ }\href@noop {} {\bibfield  {journal} {\bibinfo  {journal}
  {The European Physical Journal B}\ }\textbf {\bibinfo {volume} {88}},\
  \bibinfo {pages} {1} (\bibinfo {year} {2015})}\BibitemShut {NoStop}%
\bibitem [{\citenamefont
  {Magnus}(1954)}]{https://doi.org/10.1002/cpa.3160070404}%
  \BibitemOpen
  \bibfield  {author} {\bibinfo {author} {\bibfnamefont {W.}~\bibnamefont
  {Magnus}},\ }\href {\doibase https://doi.org/10.1002/cpa.3160070404}
  {\bibfield  {journal} {\bibinfo  {journal} {Communications on Pure and
  Applied Mathematics}\ }\textbf {\bibinfo {volume} {7}},\ \bibinfo {pages}
  {649} (\bibinfo {year} {1954})}\BibitemShut {NoStop}%
\bibitem [{\citenamefont {Blanes}\ \emph {et~al.}(2009)\citenamefont {Blanes},
  \citenamefont {Casas}, \citenamefont {Oteo},\ and\ \citenamefont
  {Ros}}]{BLANES2009151}%
  \BibitemOpen
  \bibfield  {author} {\bibinfo {author} {\bibfnamefont {S.}~\bibnamefont
  {Blanes}}, \bibinfo {author} {\bibfnamefont {F.}~\bibnamefont {Casas}},
  \bibinfo {author} {\bibfnamefont {J.}~\bibnamefont {Oteo}}, \ and\ \bibinfo
  {author} {\bibfnamefont {J.}~\bibnamefont {Ros}},\ }\href {\doibase
  https://doi.org/10.1016/j.physrep.2008.11.001} {\bibfield  {journal}
  {\bibinfo  {journal} {Physics Reports}\ }\textbf {\bibinfo {volume} {470}},\
  \bibinfo {pages} {151} (\bibinfo {year} {2009})}\BibitemShut {NoStop}%
\bibitem [{\citenamefont {Haeberlen}\ and\ \citenamefont
  {Waugh}(1968)}]{CoherentAver}%
  \BibitemOpen
  \bibfield  {author} {\bibinfo {author} {\bibfnamefont {U.}~\bibnamefont
  {Haeberlen}}\ and\ \bibinfo {author} {\bibfnamefont {J.~S.}\ \bibnamefont
  {Waugh}},\ }\href {\doibase 10.1103/PhysRev.175.453} {\bibfield  {journal}
  {\bibinfo  {journal} {Phys. Rev.}\ }\textbf {\bibinfo {volume} {175}},\
  \bibinfo {pages} {453} (\bibinfo {year} {1968})}\BibitemShut {NoStop}%
\bibitem [{\citenamefont {Van~Vleck}(1929)}]{PhysRev.33.467}%
  \BibitemOpen
  \bibfield  {author} {\bibinfo {author} {\bibfnamefont {J.~H.}\ \bibnamefont
  {Van~Vleck}},\ }\href {\doibase 10.1103/PhysRev.33.467} {\bibfield  {journal}
  {\bibinfo  {journal} {Phys. Rev.}\ }\textbf {\bibinfo {volume} {33}},\
  \bibinfo {pages} {467} (\bibinfo {year} {1929})}\BibitemShut {NoStop}%
\bibitem [{\citenamefont {Evans}\ and\ \citenamefont
  {Powles}(1967)}]{Evans_1967}%
  \BibitemOpen
  \bibfield  {author} {\bibinfo {author} {\bibfnamefont {W.~A.~B.}\
  \bibnamefont {Evans}}\ and\ \bibinfo {author} {\bibfnamefont {J.~G.}\
  \bibnamefont {Powles}},\ }\href {\doibase 10.1088/0370-1328/92/4/327}
  {\bibfield  {journal} {\bibinfo  {journal} {Proceedings of the Physical
  Society}\ }\textbf {\bibinfo {volume} {92}},\ \bibinfo {pages} {1046}
  (\bibinfo {year} {1967})}\BibitemShut {NoStop}%
\bibitem [{\citenamefont {Roberts}\ and\ \citenamefont
  {Wiseman}(2022{\natexlab{a}})}]{PhysRevB.105.195412}%
  \BibitemOpen
  \bibfield  {author} {\bibinfo {author} {\bibfnamefont {M.~M.}\ \bibnamefont
  {Roberts}}\ and\ \bibinfo {author} {\bibfnamefont {T.}~\bibnamefont
  {Wiseman}},\ }\href {\doibase 10.1103/PhysRevB.105.195412} {\bibfield
  {journal} {\bibinfo  {journal} {Phys. Rev. B}\ }\textbf {\bibinfo {volume}
  {105}},\ \bibinfo {pages} {195412} (\bibinfo {year}
  {2022}{\natexlab{a}})}\BibitemShut {NoStop}%
\bibitem [{\citenamefont {Iorio}\ and\ \citenamefont
  {Pais}(2022)}]{PhysRevB.106.157401}%
  \BibitemOpen
  \bibfield  {author} {\bibinfo {author} {\bibfnamefont {A.}~\bibnamefont
  {Iorio}}\ and\ \bibinfo {author} {\bibfnamefont {P.}~\bibnamefont {Pais}},\
  }\href {\doibase 10.1103/PhysRevB.106.157401} {\bibfield  {journal} {\bibinfo
   {journal} {Phys. Rev. B}\ }\textbf {\bibinfo {volume} {106}},\ \bibinfo
  {pages} {157401} (\bibinfo {year} {2022})}\BibitemShut {NoStop}%
\bibitem [{\citenamefont {Roberts}\ and\ \citenamefont
  {Wiseman}(2022{\natexlab{b}})}]{PhysRevB.106.157402}%
  \BibitemOpen
  \bibfield  {author} {\bibinfo {author} {\bibfnamefont {M.~M.}\ \bibnamefont
  {Roberts}}\ and\ \bibinfo {author} {\bibfnamefont {T.}~\bibnamefont
  {Wiseman}},\ }\href {\doibase 10.1103/PhysRevB.106.157402} {\bibfield
  {journal} {\bibinfo  {journal} {Phys. Rev. B}\ }\textbf {\bibinfo {volume}
  {106}},\ \bibinfo {pages} {157402} (\bibinfo {year}
  {2022}{\natexlab{b}})}\BibitemShut {NoStop}%
\bibitem [{\citenamefont {Oliva-Leyva}\ and\ \citenamefont
  {Naumis}(2015)}]{OLIVALEYVA20152645}%
  \BibitemOpen
  \bibfield  {author} {\bibinfo {author} {\bibfnamefont {M.}~\bibnamefont
  {Oliva-Leyva}}\ and\ \bibinfo {author} {\bibfnamefont {G.~G.}\ \bibnamefont
  {Naumis}},\ }\href {\doibase https://doi.org/10.1016/j.physleta.2015.05.039}
  {\bibfield  {journal} {\bibinfo  {journal} {Physics Letters A}\ }\textbf
  {\bibinfo {volume} {379}},\ \bibinfo {pages} {2645} (\bibinfo {year}
  {2015})}\BibitemShut {NoStop}%
\end{thebibliography}%
\bibliographystyle{apsrev4-1}
\end{document}